\documentclass[prl, final, amsmath, amssymb, superscriptaddress, showpacs, a4paper, aps, 10pt, hidelinks, numbers, sort&compress, floats, twocolumn, balancelastpage]{revtex4-1}

\usepackage{color,amsmath,amssymb,amsthm,times,graphics,graphicx,bm,bbm,dcolumn}
\usepackage{epsfig}
\usepackage{epstopdf}
\usepackage{graphicx}

\usepackage{amssymb}
\usepackage{braket}

\usepackage{bm}
\usepackage[dvipsnames]{xcolor}
\usepackage[normalem]{ulem}
\definecolor{mygray}{gray}{0.6}

\usepackage{hyperref}
\hypersetup{linkcolor=blue, citecolor=blue, urlcolor=blue}

\usepackage[utf8]{inputenc}

\usepackage[resetlabels,labeled]{multibib}
\newcites{SM}{References}

\newcommand{\1}{\ensuremath{\left|1 \right\rangle}}
\newcommand{\2}{\ensuremath{\left|2 \right\rangle}}

\definecolor{britishracinggreen}{rgb}{0.0, 0.26, 0.15}
\definecolor{bulgarianrose}{rgb}{0.28, 0.02, 0.03}
\definecolor{darkred}{rgb}{0.90,0,0}
\definecolor{darkgreen}{rgb}{0,0.60,.2}
\definecolor{darkblue}{rgb}{0,0,1}
\definecolor{orange}{cmyk}{0,0.6,0.8,0}
\definecolor{lightblue}{rgb}{0.3,0.5,1}
\definecolor{lightgreen}{rgb}{0.4,0.80,.4}

\newcommand{\Li}{$^{6}$Li }

\begin{document}
\title{
Connecting dissipation and phase slips in a Josephson junction between fermionic superfluids}
\author{A.~Burchianti}
\affiliation{Istituto Nazionale di Ottica del Consiglio Nazionale delle Ricerche (INO-CNR), 50019 Sesto Fiorentino, Italy}
\affiliation{\mbox{LENS and Dipartimento di Fisica e Astronomia, Universit\`{a} di Firenze, 50019 Sesto Fiorentino, Italy}}
\author{F. Scazza}
\email[Corresponding author. E-mail address: ]{scazza@lens.unifi.it}
\affiliation{Istituto Nazionale di Ottica del Consiglio Nazionale delle Ricerche (INO-CNR), 50019 Sesto Fiorentino, Italy}
\affiliation{\mbox{LENS and Dipartimento di Fisica e Astronomia, Universit\`{a} di Firenze, 50019 Sesto Fiorentino, Italy}}
\author{A. Amico}
\affiliation{\mbox{LENS and Dipartimento di Fisica e Astronomia, Universit\`{a} di Firenze, 50019 Sesto Fiorentino, Italy}}
\author{G. Valtolina}
\altaffiliation[Present address: ]{JILA, University of Colorado, Boulder, CO 80309, USA}
\affiliation{Istituto Nazionale di Ottica del Consiglio Nazionale delle Ricerche (INO-CNR), 50019 Sesto Fiorentino, Italy}
\affiliation{\mbox{LENS and Dipartimento di Fisica e Astronomia, Universit\`{a} di Firenze, 50019 Sesto Fiorentino, Italy}}
\author{\\J. A. Seman}
\affiliation{Instituto de Fisica, Universidad Nacional Aut\'onoma de M\'exico, 01000 Ciudad de M\'exico, Mexico}
\author{C. Fort}
\affiliation{Istituto Nazionale di Ottica del Consiglio Nazionale delle Ricerche (INO-CNR), 50019 Sesto Fiorentino, Italy}
\affiliation{\mbox{LENS and Dipartimento di Fisica e Astronomia, Universit\`{a} di Firenze, 50019 Sesto Fiorentino, Italy}}
\author{M. Zaccanti}
\affiliation{Istituto Nazionale di Ottica del Consiglio Nazionale delle Ricerche (INO-CNR), 50019 Sesto Fiorentino, Italy}
\affiliation{\mbox{LENS and Dipartimento di Fisica e Astronomia, Universit\`{a} di Firenze, 50019 Sesto Fiorentino, Italy}}
\author{M. Inguscio}
\affiliation{Istituto Nazionale di Ottica del Consiglio Nazionale delle Ricerche (INO-CNR), 50019 Sesto Fiorentino, Italy}
\affiliation{\mbox{LENS and Dipartimento di Fisica e Astronomia, Universit\`{a} di Firenze, 50019 Sesto Fiorentino, Italy}}
\author{G. Roati}
\affiliation{Istituto Nazionale di Ottica del Consiglio Nazionale delle Ricerche (INO-CNR), 50019 Sesto Fiorentino, Italy}
\affiliation{\mbox{LENS and Dipartimento di Fisica e Astronomia, Universit\`{a} di Firenze, 50019 Sesto Fiorentino, Italy}}

\begin{abstract}
We study the emergence of dissipation in an atomic Josephson junction between weakly-coupled superfluid Fermi gases. We find that vortex-induced phase slippage is the dominant microscopic source of dissipation across the BEC-BCS crossover. We explore different dynamical regimes by tuning the bias chemical potential between the two superfluid reservoirs. For small excitations, we observe dissipation and phase coherence to coexist, with a resistive current followed by well-defined Josephson oscillations. We link the junction transport properties to the phase-slippage mechanism, finding that vortex nucleation is primarily responsible for the observed trends of conductance and critical current. For large excitations, we observe the irreversible loss of coherence between the two superfluids, and transport cannot be described only within an uncorrelated phase-slip picture. Our findings open new directions for investigating the interplay between dissipative and superfluid transport in strongly-correlated Fermi systems, and general concepts in out-of-equlibrium quantum systems. 
\end{abstract}
 
\maketitle

The frictionless flow of particles in superfluids and superconductors is a direct manifestation of macroscopic quantum phase coherence.
But such systems, in certain conditions, exhibit a non-zero resistivity stemming from dissipative microscopic processes \cite{barone,Tinkham,Halperin,Var14}. 
In particular, when superfluids flow through constrictions or channels, the maximum flow is limited by the stochastic nucleation of vortices \cite{Var14,Feynman}. Vortices traversing the channel cause the phase to slip and remove energy from the superflow, that ceases to be dissipationless \cite{Var14, And66, avenel1985}. 
Phase slips represent the fundamental dissipation mechanism in superfluid helium \cite{And66, avenel1985,Var14}, and play also an important role for the resistivity of thin superconducting wires and two-dimensional films \cite{Langer66,Bezryadin,Kamenev2014,Halperin}. Understanding and controlling dissipation in superfluids is crucial for developing novel quantum devices with ultimate sensitivity \cite{SQUID, Packard2012}. In this context, ultracold atomic gases in tailored optical potentials have emerged as a powerful platform \cite{Ventra}. Dissipative dynamics has been observed in Bose-Einstein condensates \cite{Lev07,DeMarco2008,Hadzibabic_2012,Wrigh_2013,Jen2014,Tanzi2016,Eckel2016}, and in superfluid Fermi gases in the presence of either weak obstacles \cite{Miller2007,Watanabe2009,Weimer2015} or bosonic counterflow \cite{Delehaye2015,Castin2015}. 
Recently, quantum transport through weak links connecting two strongly interacting fermionic superfluids has also been realized, observing different dissipation mechanisms akin to those typical of solid-state devices \cite{Sta12, husmann2015, valtolina}. 
In particular, for a planar Josephson junction, we revealed the onset of vortex-induced dissipation upon reducing the coupling between the reservoirs \cite{valtolina}.

\begin{figure}[b!]
\center
\vspace*{0pt}
\includegraphics[width= 0.9\columnwidth]{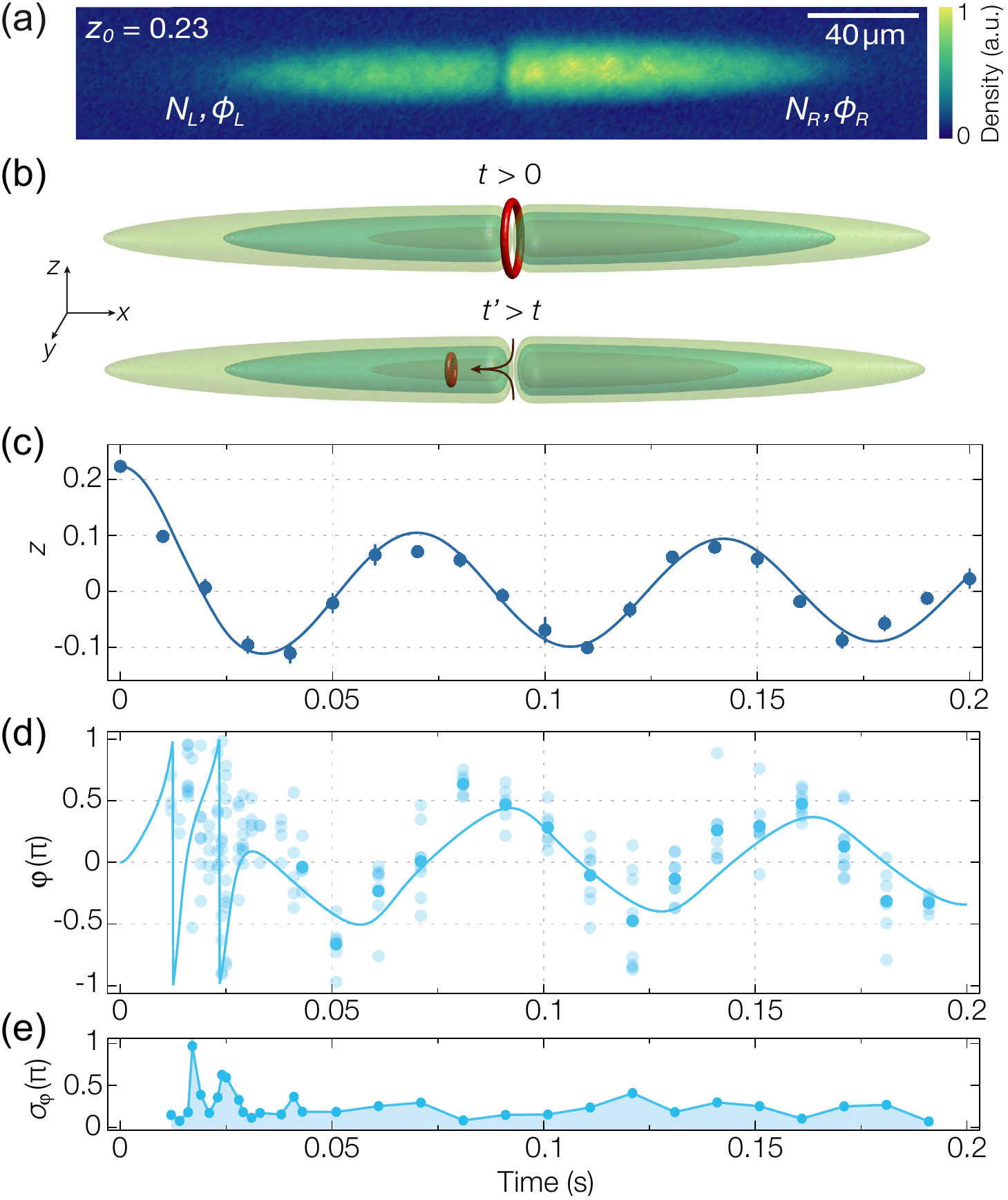}
\caption{ (a) \textit{In situ} density profile of an atomic superfluid bisected by a thin barrier with $z_0\simeq 0.23$. (b) Sketch of a phase-slip event: a vortex ring is created within the junction at time $t$, and it subsequently penetrates into the superfluid bulk after shrinking. 
(c)-(d) Typical population imbalance $z$ and relative phase $\varphi$ evolutions for a molecular BEC at $1/(k_F a) = 4.6$ and $V_0/\mu=0.7$. Both solid curves are obtained by a single fit of the measured $z(t)$ with the solution of a RSJ-like circuit model (see text). Error bars in panel (c) denote standard errors over at least five independent measurements, while light (dark) circles in panel (d) represent single (averaged) experimental realizations. (e) Standard deviation of the measured $\varphi$. The two peaks at short times are associated with stochastic phase-slip events, where shot-to-shot fluctuations are maximized.}
\label{schema}
\end{figure}

\begin{figure*}
\centering
\includegraphics[width=178mm]{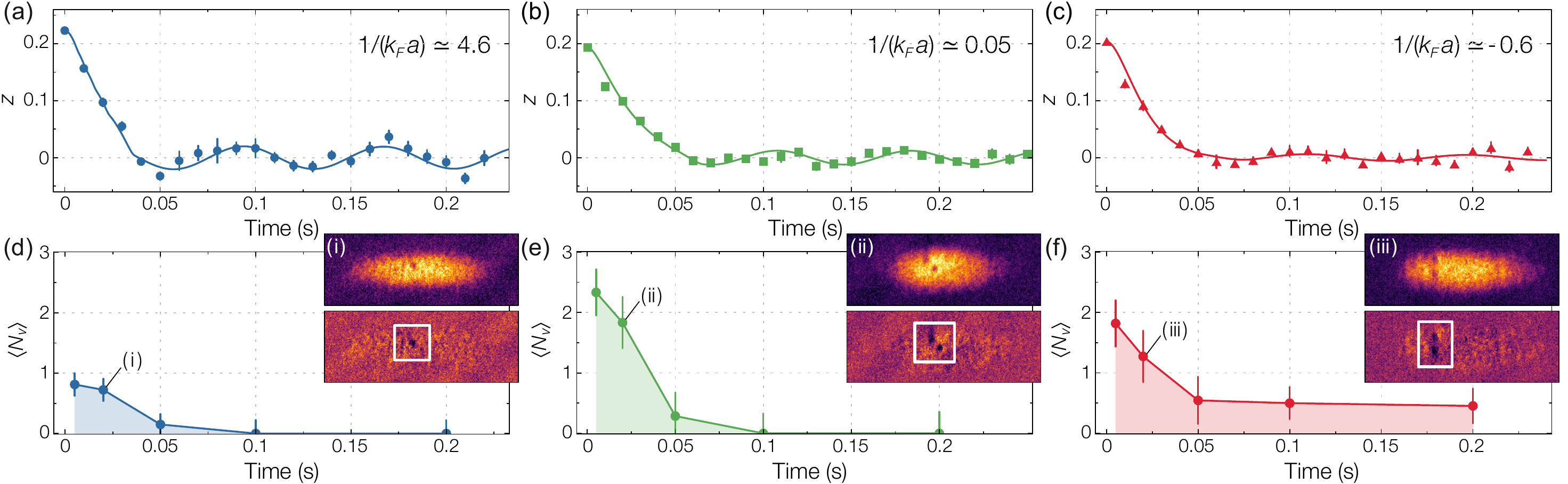}
\caption{(a)-(c) Evolution of the relative population imbalance with $z_0\simeq 0.2$ for (a) a molecular BEC at $V_0/\mu \simeq1$, (b) a unitary Fermi gas at $V_0/\mu \simeq 0.9$ and (c) a BCS superfluid at $V_0/\mu \simeq 0.9$.  Solid lines are fitted to the data with the solution of the circuital model described in the text. Error bars denote standard errors over at least five measurements. (d)-(f)  Evolution of average vortex counts $\left\langle N_v\right\rangle$ for the same experimental conditions as in panels (a)-(c). The error bars are estimated as $\sqrt{\sigma_ {Nv}^{2}+1/Z}$, with $\sigma_ {Nv}$ the standard deviation of the mean and $Z$ the number of experimental measurements. The insets (i)-(iii) show typical time-of-flight images after $20$\,ms of evolution, where vortex defects are clearly visible. %
Residual images are also displayed, obtained by subtracting the density distribution of a cloud without excitations. %
}
\label{figura2}
\end{figure*}

In this work, we demonstrate the direct connection between phase slips and dissipative transport across a Josephson junction between atomic superfluids throughout the Bose-Einstein condensate (BEC) -- Bardeen-Cooper-Schrieffer (BCS) crossover. Notwithstanding the different nature of the superfluids herein investigated, we find that phase slippage is the dominant mechanism fostering dissipation of the superfluid energy. We directly detect phase-slip events, emerging as vortex excitations created within the junction and shed into the bulk, and we show the link between the phase-slippage rate and the chemical potential difference across the junction. In the regime of low phase-slippage rate, when few excitations are nucleated, the system exhibits a transient resistive current followed by Josephson plasma oscillations. For larger initial excitations, instead, strong dissipation irreversibly suppresses the Josephson coupling across the junction. We fully characterize the junction by extracting the conductance $G$ and the critical current $I_c$ through a resistively-shunted junction model, similar to that used for ordinary superconducting junctions \cite{barone, Tinkham}. We find $G$ to depend linearly on the superfluid density in the weak-link region, whereas $I_c$ is bounded by the value of the local Landau critical velocity. Our observations deviate from the behaviour of typical superconducting junctions \cite{barone, Tinkham} or unitary superfluids connected by a quantum point contact \cite{husmann2015}, where dissipation is governed by pair-breaking effects that hinder phase coherence.

We produce fermionic superfluids of $N \simeq 10^5$ atom pairs, cooling a balanced mixture of the two lowest spin states of $^6$Li to $T/T_F \simeq 0.1$ \cite{Bur14, SM}. Here, $T_F$ is the Fermi temperature, $k_B T_F=E_F=\hbar \,(6N\omega_x \omega_y \omega_z )^{1/3}$, where $k_B$ and $\hbar\!=\!h/(2\pi)$ are the Boltzmann and reduced Planck constants, and $(\omega_x, \omega_y , \omega_z)\simeq 2\pi \times (14, 140, 160)\,$Hz are the trapping frequencies. 
Interactions are parametrized by $1/(k_F a)$, where $a$ is the $s$-wave scattering length and $k_F$ is the Fermi wave-vector defined by $E_F = \hbar^2 k_F^2/(2m)$, with $m$ being $^6$Li atomic mass. The scattering length between the two spin states is adjusted via a broad Feshbach resonance located at 832\,G \cite{zurn2013}. Hereafter, we focus on three distinct regimes of superfluidity: (i) a molecular BEC at $1/(k_Fa) \simeq 4.6$, (ii) a unitary superfuid at $1/(k_Fa) \simeq 0.05$, and (iii) a BCS superfluid at $1/(k_Fa) \simeq -0.6$. 
We realize the atomic Josephson junction by splitting the superfluid into two weakly coupled reservoirs using a thin Gaussian optical barrier of variable height $V_0$ and 1/$e^2$ width $w\simeq 2\,\mu$m, few times wider than the superfluid coherence length across the BEC-BCS crossover \cite{valtolina,SM}. 
The dynamics is described by the relative population imbalance $z = (N_R - N_L)/N$, corresponding to a chemical potential difference $\Delta\mu =  \mu_R-\mu_L$ across the junction, and the relative phase $\varphi = \varphi_L-\varphi_R$ between the two reservoirs, where $N_R$ ($N_L$) and $\varphi_R$ ($\varphi_L$) are the pair population and phase of the right (left) reservoir, respectively (see Fig.~\ref{schema}(a)). 
Experimentally, we monitor the relative population imbalance and phase evolutions by absorption imaging of the in-situ density and of the interfering reservoirs during a time-of-flight expansion from the trap, respectively  \cite{valtolina}.

For barrier strengths $V_0 \sim \mu$, the system dynamics is determined essentially by the competition between the Josephson tunnelling and charging energy \cite{Sme97, Zapata, Meier2001}. When the tunnelling dominates, for small initial excitations, $z$ and $\varphi$ undergo Josephson plasma oscillations. In the opposite limit of large $\Delta\mu_0$ and in the absence of dissipation, the atomic system is expected to enter the macroscopic self-trapping state (MQST), where a linear increase of $\varphi$ drives small-amplitude oscillations of $z$ around a non-zero value at a frequency $\sim \Delta\mu_0/\hbar$ \cite{Sme97, Zapata, Alb05, Abbarchi2013, Zou2014, Fattori2016}. To explore the latter regime, we prepare a tunable initial imbalance $z_{0}$, corresponding to $\Delta\mu_0/\mu \leq 0.4$, with $\mu$ denoting the bulk chemical potential at equilibrium. 
By lowering the barrier height to the target value $V_0$ at time $t=0$, we induce a current $I=\dot{k}$, where $k = zN/2$ (see Fig.~\ref{schema}(a) and \cite{SM} for details). In Fig.~\ref{schema}(c)-(d), $z(t)$ and $\varphi(t)$ are displayed for a molecular BEC with $z_0 \simeq 0.23$ and $V_0/ \mu \simeq 0.7$. We observe that $z$ displays an initial decay alongside a fast variation of $\varphi$ in the range $\left(-\pi,\pi\right)$. Thereafter, both $z(t)$ and $\varphi(t)$ oscillate around zero at the Josephson plasma frequency $\omega<\omega_x$ with a relative phase shift of about $\pi/2$. A similar behaviour is observed in all explored regimes of superfluidity, as shown in Fig.~\ref{figura2}(a)-(c), where $z(t)$ is compared for BEC, unitary and BCS superfluids with $z_0 \simeq 0.2$ and $V_0/ \mu \simeq 1$. 
While the observed initial variation of $\varphi$ is consistent with a running-phase evolution, the irreversible decay of $z$ reflects the instability of MQST \cite{Zapata, Ruostekoski1998, Meier2001, Zou2014}. This highlights the presence of dissipation mechanisms, that could stem from either thermal \cite{Lev07} or collective excitations \cite{valtolina}, which however do not destroy the coherent coupling across the junction, as demonstrated by the Josephson dynamics emerging after dissipation.
The combination of running-phase evolution and dissipative flow, closely resembling the dynamics of strongly coupled superfluid ${}^4$He reservoirs at $T < T_\lambda$ \cite{Packard2012}, suggests the occurrence of stochastic phase-slip events (see Fig.~\ref{schema}(b)). This is also supported by the significant fluctuations of $\varphi$ detected at short times (see~Fig.~\ref{schema}(e)).  

We gain further insight into the microscopic origin of dissipation by monitoring the atomic cloud in time-of-flight after adiabatically removing the barrier \cite{valtolina,SM}. We observe the initial drop of $z$ to be accompanied by the presence of vortex defects in the superfluid bulk, detected as local density depletions predominantly located within the reservoir at lower initial chemical potential (see insets of Fig.~\ref{figura2}(d)-(f)). 
In Fig.~\ref{figura2}(d)-(f), we show the time evolution of the mean vortex count $\left\langle N_v\right\rangle$ extracted from typically 15 time-of-flight images, acquired in the same experimental conditions as in Fig.~\ref{figura2}(a)-(c). In all explored interaction regimes, $\left\langle N_v\right\rangle$ is found to decay within the same timescale as $z$. Such correlated trend of $z(t)$ and $\left\langle N_v\right\rangle\!(t)$ strongly supports the scenario of dissipation driven by vortex-induced phase-slip events, where the vortex nucleation rate $\gamma$, i.e.~the phase-slip rate, follows the Josephson-Anderson relation, $\gamma \simeq \dot{\varphi}/(2\pi)  = \Delta\mu/h$ \cite{And66,avenel1985}. Accordingly, for a given $z_0$, $\left\langle N_v\right\rangle$ becomes larger when moving from the BEC to the crossover regime, reflecting the increase of $\Delta\mu_0$. %
Once $z(t)$ has dropped below a critical value, the vortex nucleation rate is strongly reduced, so that $\left\langle N_v\right\rangle$ remains low and pure Josephson dynamics is established.   
$\left\langle N_v\right\rangle$ is also determined by the vortex lifetime, which depends on the interaction strength and is limited by vortex decay into sound-like excitations, favored by the density kink at the barrier position \cite{Suthar, SM, proukakis}. Although sound waves must ultimately dissipate into heat, we do not observe any related appreciable reduction of the condensed fraction within the measurement timescale \cite{SM}. 

Our observations agree with simulations of weakly-linked three-dimensional bosonic superfluids, showing that phase slippage typically arise from vortex rings nucleated within the barrier at the edge of the atomic cloud, which shrink and cross the junction region perpendicularly to the flow (see Fig.~\ref{schema}(b)) \cite{piazza, PiazzaJOPB, Abad}. We confirm this scenario by performing numerical simulations in the BEC and unitary regimes with the zero-temperature extended Thomas-Fermi model (ETFM), based on a generalized Gross-Pitaevskii equation for the pairs wavefunction including the chemical potential from Quantum Monte-Carlo calculations across the entire crossover \cite{Forbes_2014, SM}. The simulations correctly capture the decay of $z$ due to vortex shedding into the bulk, 
which is favored by the multimode character of our junction \cite{SM,proukakis}. Experimentally, we observe defects predominantly oriented along the tighter confining trap axis, i.e.~the imaging line-of-sight (see Fig.~\ref{figura2}(d)-(f)). This is consistent with the instability of vortex rings towards breaking up into vortex lines in a radially asymmetric trap with $\omega_y<\omega_z$ \cite{MIT_solitons}, assisted also by the slow barrier removal prior to imaging \cite{SM, proukakis}.  

\begin{figure}[t!]
\center
\includegraphics[width= \columnwidth]{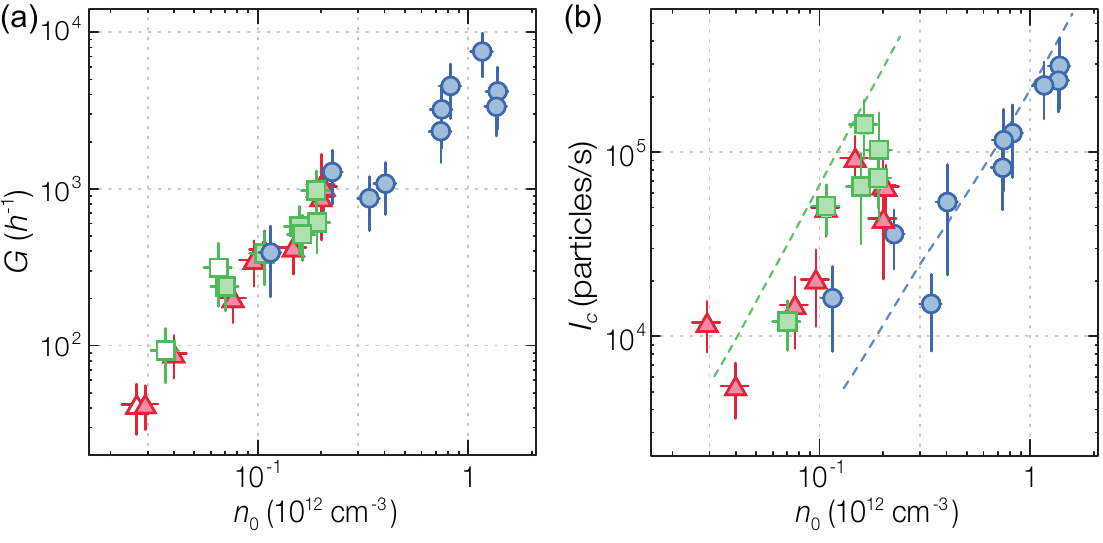}
\caption{(a) Conductance $G$ and (b) critical current $I_c$ as a function of the central pair-density $n_0$ inside the barrier for BEC (blue circles), unitary (green squares) and BCS (red triangles) superfluids. For filled (empty) symbols, $G$ is obtained through the RSJ-like (RC) circuit model (see text). The dashed lines in panel (b) represent the calculated upper bound $I_{c0}$ \cite{SM}, shown for comparison for BEC (blue) and unitary (green) superfluids. In both panels, horizontal error bars account for the typical $20\%$ uncertainty on atom number, while vertical ones combine this with fitting standard errors.}
\label{figura3}
\vspace*{0pt}
\end{figure}
\begin{figure}[t!]
\centering
\includegraphics[width=\columnwidth]{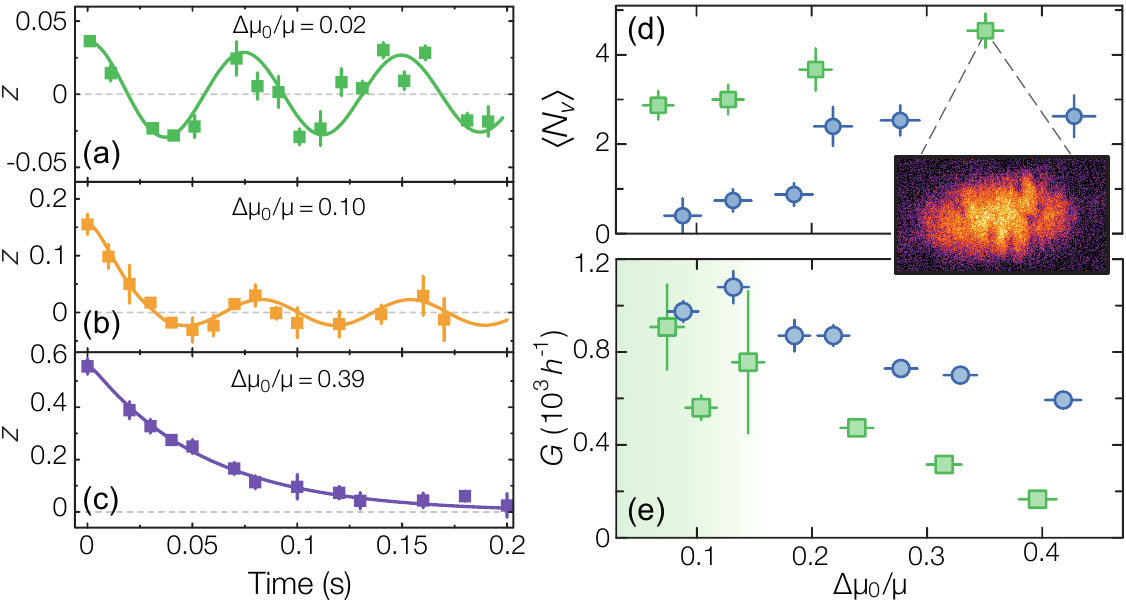}
\caption{(a)-(c) Crossover from Josephson to purely dissipative dynamics in a unitary superfluid at $V_0/\mu\simeq0.9$. The initial bias potentials $\Delta\mu_0/\mu$ are: (a) 0.02, (b) 0.10 and (c) 0.39. (d) Average vortex counts $\left\langle N_v\right\rangle$ and (e) conductance $G$ as a function of $\Delta\mu_0/\mu$ for BEC superfluids at $V_0/\mu\simeq1.3$ (blue circles) and unitary superfluids at $V_0/\mu\simeq0.9$ (green squares). The central density $n_0$ is about three times larger for BECs with respect to unitary gases. The green shaded region indicates the range of $\Delta\mu_0/\mu$ where $I_c>0$ at unitarity. Vertical error bars are computed as described in the caption of Fig.~\ref{figura2}~and~\ref{figura3}, while horizontal ones result from the experimental uncertainty on $z_0$, that is typically of $\pm 2\%$. Inset: time-of-flight image of an expanding unitary superfluid for $\Delta\mu_0/\mu\simeq 0.35$, where several vortex defects are visible.}
\label{figura4}
\vspace*{0pt}
\end{figure}

To quantitatively characterize the transport through the junction, we model it with an equivalent circuit made of three parallel elements: a Josephson weak link with a current-phase relation \mbox{$I_J=-I_c\sin(\varphi)$}, a shunt resistance $R$ and a series $LC$ \cite{SM, Eckel2016}. In this way, we extract the conductance $G=R^{-1}$ and critical current $I_c$. The use of this model is justified, as we find the dissipative current to be ohmic with a linear current-bias relation \cite{SM}. %
This approach is equivalent to the resistively-shunted junction (RSJ) model \cite{barone,Stewart68,McCumber68}, %
and it incorporates both a Josephson and a resistive current $I_R = -G \Delta\mu$, where the resistance $R$ can account for various dissipation mechanisms.
For superconducting junctions, these typically involve the breaking of Cooper pairs \cite{barone, Giaever}. Here instead, we argue that resistivity originates from vortex-induced phase slippage rather than from unpaired fermions.  
The measured $z(t)$ is well fitted by the numerical solution of the model, where $R$ and $I_c$ are left as free parameters (see solid lines in Fig.~\ref{schema},~\ref{figura3} and~\ref{figura4}) \cite{SM}. 
For initial bias potentials $\Delta \mu_0/\mu$ between 0.05 and 0.2, and barrier heights ranging from $V_0/\mu\sim 0.6$ to 1.5, we find that $G$ and $I_c$ do not depend appreciably upon $\Delta \mu_0$ at a given $V_0/\mu$. 
This is expected for phase-slip-driven dissipation \cite{avenel1985,Jen2014}, in a regime where only few, uncorrelated topological excitations are nucleated into the superfluid (see Fig.~\ref{figura2}(d)-(f)).
For the largest values of $V_0$, where $I_c$ is strongly reduced \cite{valtolina} and Josephson oscillations are not experimentally resolved, $G$ is extracted using a simple RC circuit model.
To directly connect the measured conductances with the phase-slippage mechanism, we express the resistive current as $I_R \propto N_{ex}\gamma$, where $N_{ex}$ is the number of particles participating to each excitation \cite{Jen2014}. %
For phase slippage, $\gamma \simeq \Delta\mu/h$, that yields $G=-I_R/\Delta\mu \propto N_{ex}/h$. Therefore, we expect $G\propto n_0$, where $n_0$ is the central density inside the barrier where the excitations form.
Even though we are not able to directly measure $n_0$, due to the $1.5\,\mu$m imaging resolution and to light-induced atom diffusion during the imaging pulse, we efficiently estimate $n_0 = n_0(V_0/\mu)$ by the equilibrium solution of the ETFM \cite {SM}. In this way, we can relate the values $G$ and $I_c$ extracted \mbox{for each value of $V_0/ \mu$ to $n_0$}. 

Figure~\ref{figura3}(a) displays the conductance $G$ in units of $h^{-1}$ as a function of $n_0$ for BEC, unitary and BCS superfluids. 
To test the linearity of the measured $G$, we fit the experimental data with a power law, $G\propto n_0^\alpha$. For BEC and unitary regimes, we indeed find $\alpha= 1.0(3)$ and $\alpha=1.1(2)$, respectively. For BCS superfluids, we instead obtain $\alpha =1.5(2)$. This non-linearity may stem from the limited accuracy of our $n_0$ estimate in the BCS regime and from additional dissipation sources such as single-particle excitations. %
Importantly, we observe approximately matching conductances at fixed $n_0$ regardless of the specific nature of the superfluid, evidencing the dominant role of phase slippage.  
The large values of $G$ highlight the composite bosonic nature of the tunnelling particles carrying the current. Furthermore, our findings elucidate the origin of the finite resistance measured for unitary superfluids connected via a quasi-two-dimensional channel \cite{Sta12}. 
In Fig.~\ref{figura3}(b), the extracted critical current $I_c$ is presented as a function of $n_0$ in the different interaction regimes.  
In contrast to $G$, we find that $I_c$ depends on the nature of the condensate across the BEC-BCS crossover. 
Resonant superfluids display the largest $I_c$ at a given $n_0$, confirming their enhanced robustness \cite{Miller2007,Weimer2015, valtolina}, also in the presence of dissipation.
$I_c$ is expected to be associated with the critical velocity for vortex nucleation at the superfluid surface inside the barrier \cite{Watanabe2009, Spuntarelli}. For BEC and unitary superfluids, the latter is predicted to be lower than the local sound speed $c$ \cite{Watanabe2009, PiazzaJOPB, Spuntarelli}, yielding an upper bound $I_{c0} = c \,n_{0x}$, where $n_{0x}$ is the radially-integrated central density \cite{SM}.
The experimental data approach the calculated $I_{c0}$, with trends in qualitative agreement (see dashed lines in Fig.~\ref{figura3}(b)). Even though $c$ increases moving towards the BCS limit, the measured $I_c$ for BCS superfluids is not larger than at unitarity, evidencing the decrease of the (Landau) critical velocity for vortex nucleation, which becomes bounded by the fermionic excitation branch \cite{Spuntarelli, Zou2014}. This is consistent with the drop of Josephson energy $E_J \propto I_c$ observed for a BCS superfluid in the tunnelling regime, where such effect is associated with condensate depletion \cite{valtolina}.
%

%
%

For $\Delta \mu_0/ \mu\gtrsim 0.2$ the junction enters a qualitatively different regime, where transport properties depend on $\Delta \mu_0$. In Fig.\ref{figura4}(a)-(c) typical evolutions $z(t)$ are shown for unitary gases at three different values of $\Delta \mu_0/ \mu$ with $V_0/\mu\simeq 0.9$. By increasing $\Delta \mu_0/\mu$, we observe the gradual loss of the Josephson oscillation visibility, with the onset of purely dissipative transport around $\Delta \mu_0/\mu \approx 0.2$. 
We connect the resistance with vortex nucleation by measuring $\left\langle N_v\right\rangle (t)$ at varying $\Delta\mu_0/\mu$, for a unitary Fermi gas and BEC at $V_0/\mu\simeq 0.9$ and $V_0/\mu\simeq 1.3$, respectively. The results are displayed in Fig.~\ref{figura4}(d). In Fig.~\ref{figura4}(d) we present also the measured $G$ as a function of $\Delta\mu_0/\mu$. In both cases, $G$ decreases for $\Delta\mu_0/\mu \gtrsim 0.2$.  
The increase of the bias potential leads to the increase of $\gamma$ and therefore of $\left\langle N_v\right\rangle$. However, the decrease of $G$ is unexpected in a linear, uncorrelated phase-slip picture, where $\dot{\varphi}\propto \Delta\mu$: the observed behavior implies that our system cannot support an arbitrary large nucleation rate. Moreover, the disappearance of Josephson oscillations suggests that the coherent coupling between the reservoirs is irreversibly affected by phase-slip proliferation \cite{Var14}. 
The presence of several vortices interacting nearby the barrier may create a local turbulent pad region \cite{Tsatsos, Bulgac}, where the superfluid density is locally suppressed. On the other hand, the accumulation of vortices may locally scramble the relative phase, thereby suppressing the critical current akin to thermal fluctuations in superconducting junctions \cite{Tinkham}. The saturation of the vortex production rate may arise from vortex reconnections and interactions \cite{trentoPRX, Bulgac}. Our observations cannot be ascribed to an increase of the sample temperature, since the condensed fraction in the BEC regime remains above 0.7, limited only by the intrinsic lifetime of the gas \cite{SM}. 

In conclusion, our findings extend the vortex-induced phase-slippage picture of dissipation typical of liquid ${}^4$He to weakly coupled, strongly correlated atomic Fermi gases. We have found that in BEC-BCS crossover superfluids phase coherence can coexist with dissipation, afforded by topological excitations that, not depleting the condensate, do not cause the breakdown of Josephson dynamics. %
Future experiments will further explore the far-from-equilibrium regime, addressing the role of vortex proliferation and mutual interactions. Moreover, it will be interesting to investigate the effect of fluctuations around the superfluid critical temperature \cite{Zhai}. Our system offers a promising platform for exploring dissipative fermionic transport phenomena like quantum turbulence \cite{Bulgac} and dissipation-driven quantum phase transitions \cite{Kamenev2014, caldeira, fisher}. %

\vspace*{2pt}
\begin{acknowledgements}
\textbf{Acknowledgements} -- We acknowledge inspiring discussions with F. Dalfovo, T. Giamarchi, F. Piazza, N. Proukakis, A. Smerzi, A. Trombettoni and  K. Xhani. Special acknowledgments to the LENS Quantum Gases group. This work was supported by the ERC through Grant No. 307032 QuFerm2D and by the Marie Sk\l{}odowska-Curie programme (fellowship to F.S.).
J.A.S. acknowledges supporting grants from UNAM-DGAPA/PAPIIT IA101716 and \mbox{CONACyT} LN-271322.
\end{acknowledgements}

\bigskip
\bigskip
\bigskip
\bigskip

\noindent{\hspace*{5mm}\small$^*$ Corresponding author. E-mail: \href{mailto:scazza@lens.unifi.it}{scazza@lens.unifi.it}}\\
\noindent{\hspace*{5mm}\small$^\dagger$ Present address: JILA, University of Colorado, Boulder, CO\\ \hspace*{7mm}80309, USA}\\

\onecolumngrid

\begin{center}
\newpage
\textbf{
Supplemental Material\\[4mm]
\large Connecting dissipation and phase slips in a Josephson junction between fermionic superfluids}\\
\vspace{4mm}
{A.~Burchianti,$^{1,2}$
F.~Scazza,$^{1,2,*}$ 
A.~Amico,$^{2}$
G.~Valtolina,$^{1,2,\dagger}$\\
J.~A.~Seman,$^{3}$
C.~Fort,$^{1,2}$
M.~Zaccanti,$^{1,2}$
M.~Inguscio,$^{1,2}$
and G.~Roati$^{1,2}$}\\
\vspace{2mm}
{\em \small
$^1$Istituto Nazionale di Ottica del Consiglio Nazionale delle Ricerche (INO-CNR), 50019 Sesto Fiorentino, Italy\\
$^2$\mbox{LENS and Dipartimento di Fisica e Astronomia, Universit\`{a} di Firenze, 50019 Sesto Fiorentino, Italy}\\
$^3$\mbox{Instituto de Fisica, Universidad Nacional Aut\'onoma de M\'exico, 01000 Ciudad de M\'exico, Mexico}\\[2mm]}
{\small$^*$ Corresponding author. E-mail: scazza@lens.unifi.it}\\
{\small$^\dagger$ Present address: JILA, University of Colorado, Boulder, CO 80309, USA}\\
\end{center}

\bigskip

\setcounter{equation}{0}
\setcounter{figure}{0}
\setcounter{table}{0}
\setcounter{section}{0}
\makeatletter
\renewcommand{\theequation}{S.\arabic{equation}}
\renewcommand{\thefigure}{S\arabic{figure}}
\renewcommand{\thetable}{S\arabic{table}}
\renewcommand{\thesection}{S.\arabic{section}}

\twocolumngrid
\setlength{\belowcaptionskip}{0pt}
\flushbottom

\section{Experimental methods}
\subsection{Sample preparation}
\begin{figure}[b!]
\center
\includegraphics[width= \columnwidth]{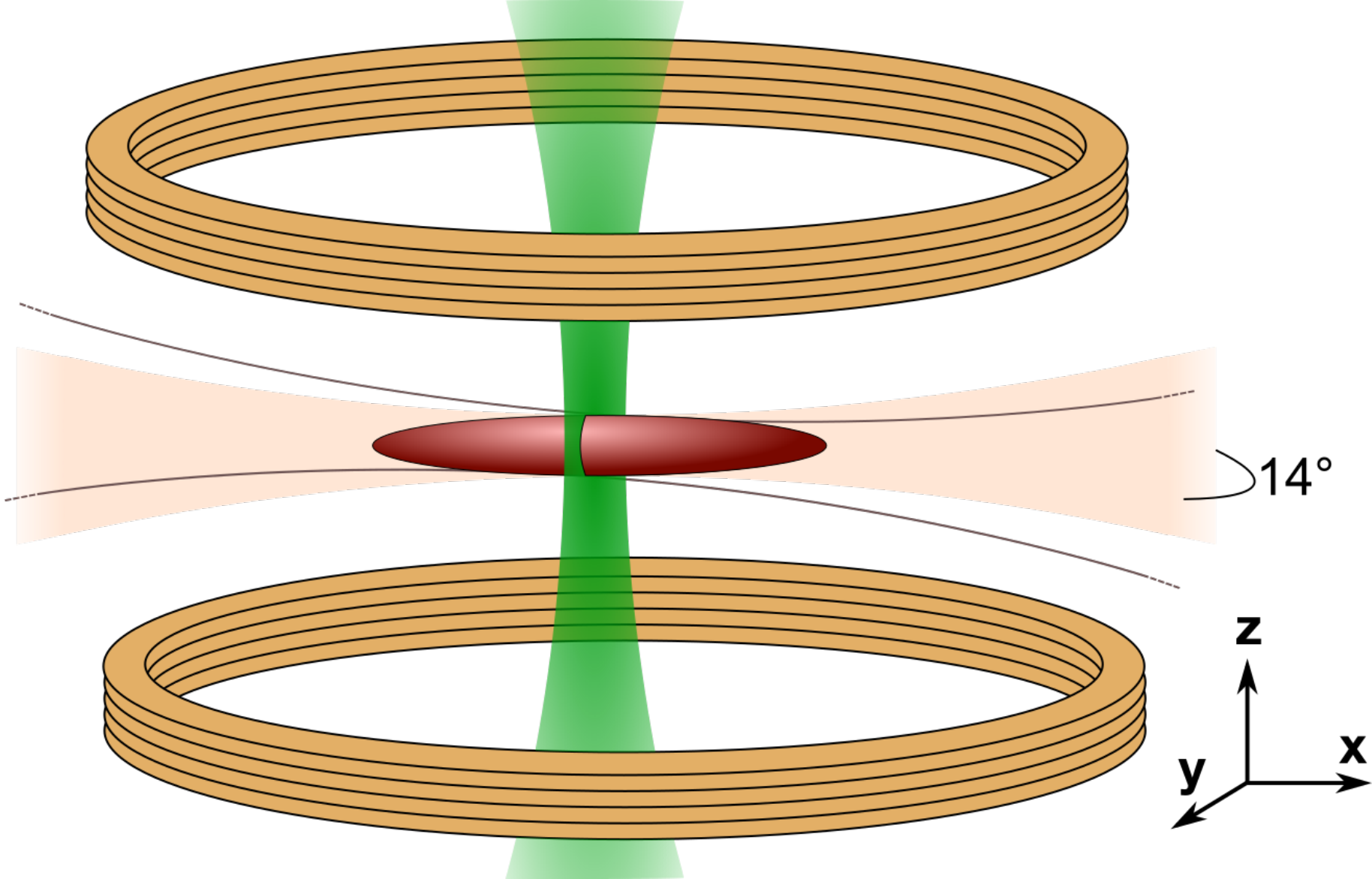}
\caption{Sketch of the experimental setup. The crossed optical dipole trap is formed by two laser beams crossing in the $xy$-plane with an angle of $14^\circ$. The focus position of one of the beams can be precisely adjusted over a range of approximately $200\mu$m by tuning the radio-frequency (RF) used to drive an acusto-optical modulator (AOM). To create a Josephson  junction, a repulsive sheet of light (depicted in green) is shone along the $z$-axis, bisecting the atomic cloud into two initially disconnected reservoirs.}
\label{setup}
\end{figure}

\begin{figure}[t]
\center
\includegraphics[width= \columnwidth]{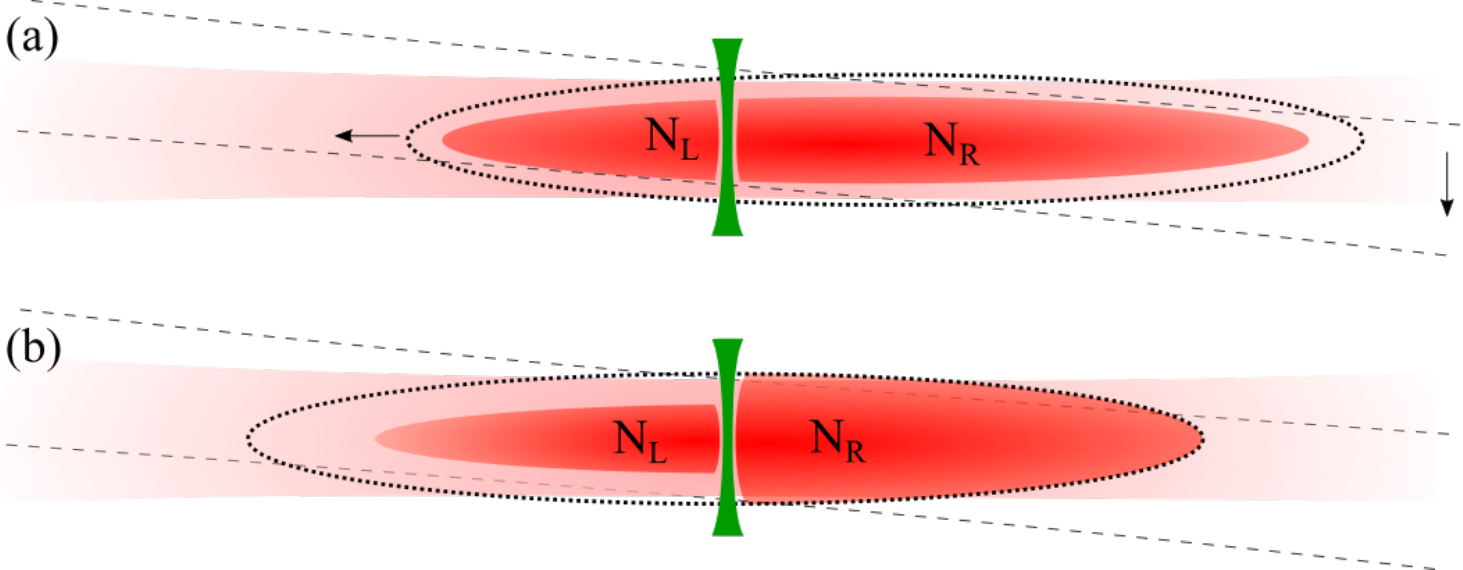}
\caption{Procedure for the preparation of the initial population imbalance. (a) In order to create the initial population imbalance between the two atomic reservoirs, the gas is evaporated in a crossed dipole trap whose center is displaced with respect to the tunnelling barrier position. (b) Once the crossed dipole trap center has been shifted to match the barrier position, yielding a symmetric trapping potential with the desired population imbalance, the evolution is started by rapidly lowering the barrier to the target height $V_0$.}
\label{imbalance}
\end{figure}

Fermionic superfluids are produced by evaporating a two-component mixture of the lowest hyperfine states of \Li in a crossed optical dipole trap. We employ the $\ket{F = 1/2, m_F = \pm1/2}$ states, labeled as \1 and \2. Following the procedure described in Refs.~\cite{Burchianti2014, valtolina}, the atomic sample is evaporatively cooled at the \1-\2 Feshbach scattering resonance located at a magnetic field of approximately 832\,G. In this way, we obtain superfluid samples of $N \approx 10^5$ atoms per spin state \cite{Burchianti2014}. At the end of the evaporation, the magnetic field is adiabatically ramped to the desired value, allowing to fine tune the inter-atomic scattering length $a$, evaluated using the magnetic-field dependence $a(B)$ reported in Ref.~\cite{Zurn2013}. The optical dipole trap is formed by two laser beams crossing horizontally with an angle of $14^\circ$ (see Fig.~\ref{setup}): the primary beam has a wavelength $\lambda_1=1064$\,nm and a beam waist $w_1\simeq45$\,$\mu$m, while the secondary beam has a wavelength $\lambda_2=1070$\,nm and it is elliptic with beam waists $w_2\simeq45$\,$\mu$m and $w'_2\simeq100$\,$\mu$m. The position of the secondary trapping beam can be finely adjusted by tuning the radio-frequency driving an acousto-optic modulator (AOM), allowing to displace the centre of the total trapping potential along the axial $x$-direction (see Fig.~\ref{imbalance}). The magnetic curvature of the Feshbach coils provides an additional weak confinement along both the $x$- and $y$-axis and a weak anti-confinement along the $z$-axis. The overall harmonic potential is characterised by radial frequencies $\omega_z\simeq160\,$Hz and $\omega_y \simeq 140\,$Hz, and an axial frequency $\omega_x \approx 14\,$Hz. Since the magnetic contribution is not fixed but depends on the magnitude of the Feshbach field, spanning from the BEC ($\sim$700\,G) to the BCS regime ($\sim$875\,G), the total value of the axial confinement frequency varies by about 10\%.

The tunnelling barrier is created as described in detail in Ref.~\cite{valtolina}. An anisotropic laser beam at 532\,nm propagating along the $z$-axis is focused on the atomic sample by using a single aspheric lens (NA $\simeq0.6$). This produces a repulsive Gaussian sheet of light that bisects the trapped atomic cloud, with a waist of $w=2.0(2)\,\mu$m and $w'=840(30)\,\mu$m along the $x$- and $y$-axis, respectively. The barrier width $w$ is only a few times larger than the superfluid coherence length across the BEC-BCS crossover and has been characterized by studying the cloud \textit{in situ} density profile, yielding results consistent with numerical simulations of the coherent oscillatory dynamics between the two reservoirs, where the barrier width is set to a fixed value (see Ref.~\cite{valtolina}). Since the barrier is almost homogeneous along the radial directions on the scale of the atomic sample, the total trapping potential acting on the pairs can be approximated as: 
\begin{equation}
\label{trapping_potential}
V(\textbf{r})=\frac{1}{2} M(\omega_x^2 x^2+\omega_y^2 y^2+\omega_z^2 z^2) +V_0\, e^{-2x^2/w^2}
\end{equation}
where $M=2m$ is the mass of an atomic pair and $V_0$ is the barrier height.

\subsection{Preparation of the initial population imbalance} 
\begin{figure}[t!]
\center
\includegraphics[width=\columnwidth]{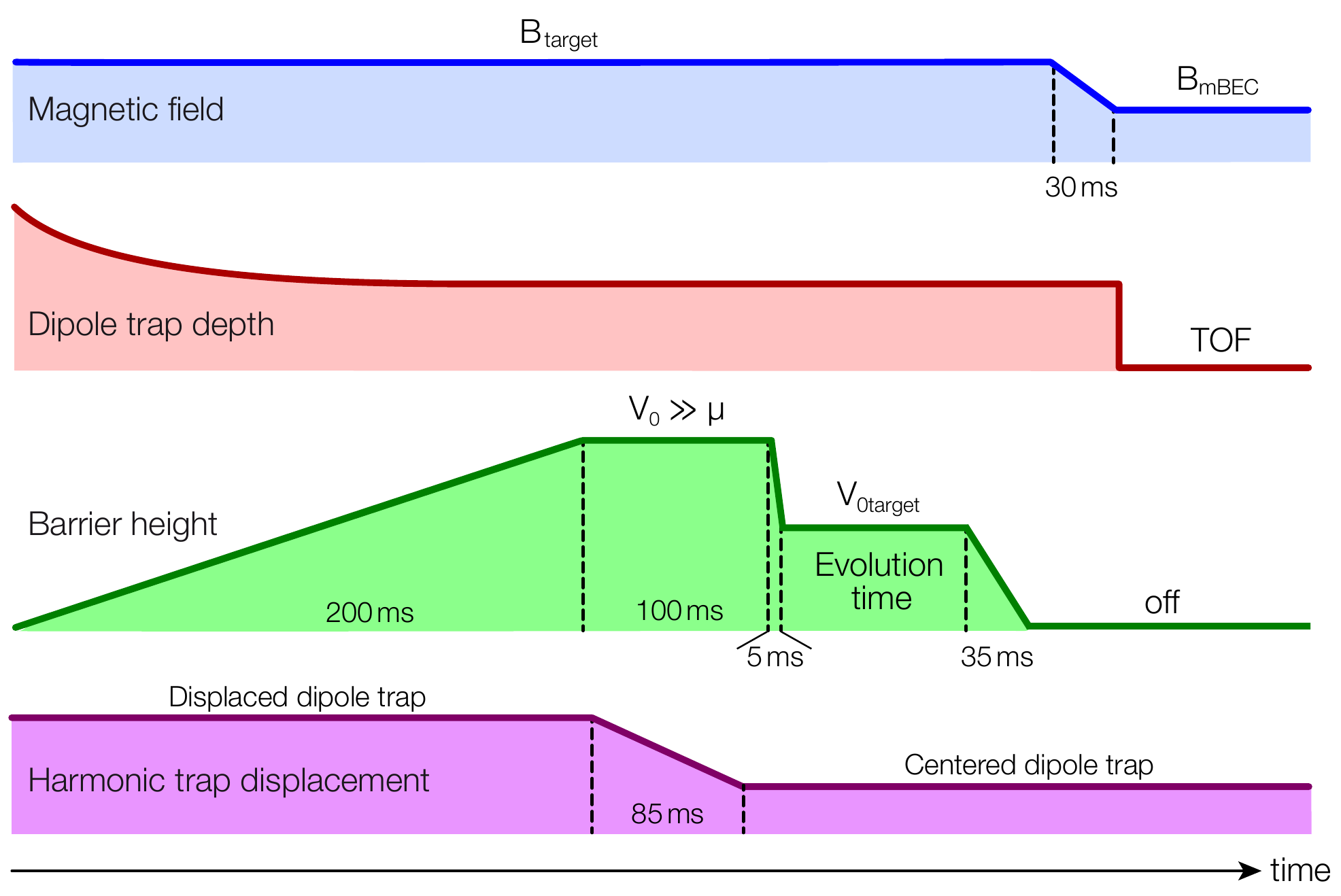}
\caption{Sketch of the experimental sequence. The Feshbach magnetic field (blue) is set to $B_\textrm{target}$ during the preparation and the evolution time, corresponding to different regimes of superfluidity across the BEC-BCS crossover. To enhance the visibility of vortex excitations, the field is ramped to $B_\textrm{mBEC}$ on the BEC side of the Feshbach resonance before imaging. The dipole trap potential (red) is exponentially ramped down during the evaporation and then kept constant. The potential barrier (green) heigth is ramped up in 200\,ms at the value $V_{0\textrm{high}} \gg \mu$ during the population imbalance preparation. Subsequently, it is rapidly ramped down to the chosen value $V_{0\textrm{target}}$. After the variable evolution time, either (i) the \textit{in situ} profile is acquired for monitoring the imbalance dynamics, or (ii) the barrier is turned completely off in $35$\,ms and the cloud is released for imaging vortices.}
\label{ramps}
\end{figure}

As described in the main text, the system dynamics is triggered by creating a non-zero initial population imbalance $z_0 = (N_R-N_L)/(N_R+N_L)$, where $N_R$ and $N_L$ are the number of pairs on the right and on the left reservoir, respectively. In order to prepare the system with $0 < z_0 \leq 0.5$, we follow the procedure depicted in Figs.~\ref{imbalance}-\ref{ramps}. First, after a superfluid has formed through evaporative cooling, we adiabatically raise the optical barrier, keeping the center of the harmonic trap conveniently shifted with respect to the barrier position (Fig. \ref{imbalance}a). Subsequently, by finely adjusting the horizontal position of the focus of one of the trapping beams, the harmonic trap center is superimposed to the barrier position to obtain an overall symmetric double-well potential (Fig. \ref{imbalance}b).  During this procedure the height of the barrier potential $V_0$ is kept well above the value of the gas chemical potential $\mu$, so as to completely suppress particle tunnelling and preserve the desired target imbalance between the two reservoirs. The value of the initial imbalance $z_0$ can be controlled by varying the initial relative displacement of the harmonic trap center. Finally, the inter-reservoir dynamics is started by rapidly lowering $V_0$ to the target value in 5\,ms (see Fig.~\ref{ramps}).

\subsection{Experimental protocol for the imaging of vortex-defects} 
To detect vortex defects in the cloud at a given evolution time during the dynamics, we release the cloud from the trap and image it after a short time-of-flight expansion. The experimental sequence is sketched in Fig.~\ref{ramps}. After initially preparing a population imbalance $z_0$, the system evolves for a variable time in the symmetric potential with a target barrier height $V_{0\textrm{target}}$. Subsequently, the barrier is adiabatically ramped down in $35$\,ms, and the magnetic field is ramped to the BEC side of the Feshbach resonance at $B_\textrm{mBEC} \simeq 700$\,G. The visibility of vortices in crossover superfluids is strongly reduced by the sharp decrease of the condensed fraction while approaching the BCS limit: the slow sweep of the Feshbach magnetic field to the BEC side of the resonance allows to convert all fermionic pairs into tightly bound molecules, emptying out the vortex cores and enhancing the defect visibility \cite{MIT_solitons}. At this point, the trapping potential is turned off. The expanding sample is then detected through high-intensity absorption imaging after a time-of-flight of $3$-$4$\,ms.

\subsection{Evolution of the condensed fraction} 
The phase-slippage mechanism removes energy from the superflow, which will eventually be dissipated into heat. For our confined reservoirs, vortex excitations are coupled to collective sound excitations, into which they can decay to subsequently dissipate into thermal excitations. 
\begin{figure}[t!]
\center
\vspace*{-8pt}
\includegraphics[width= 0.95\columnwidth]{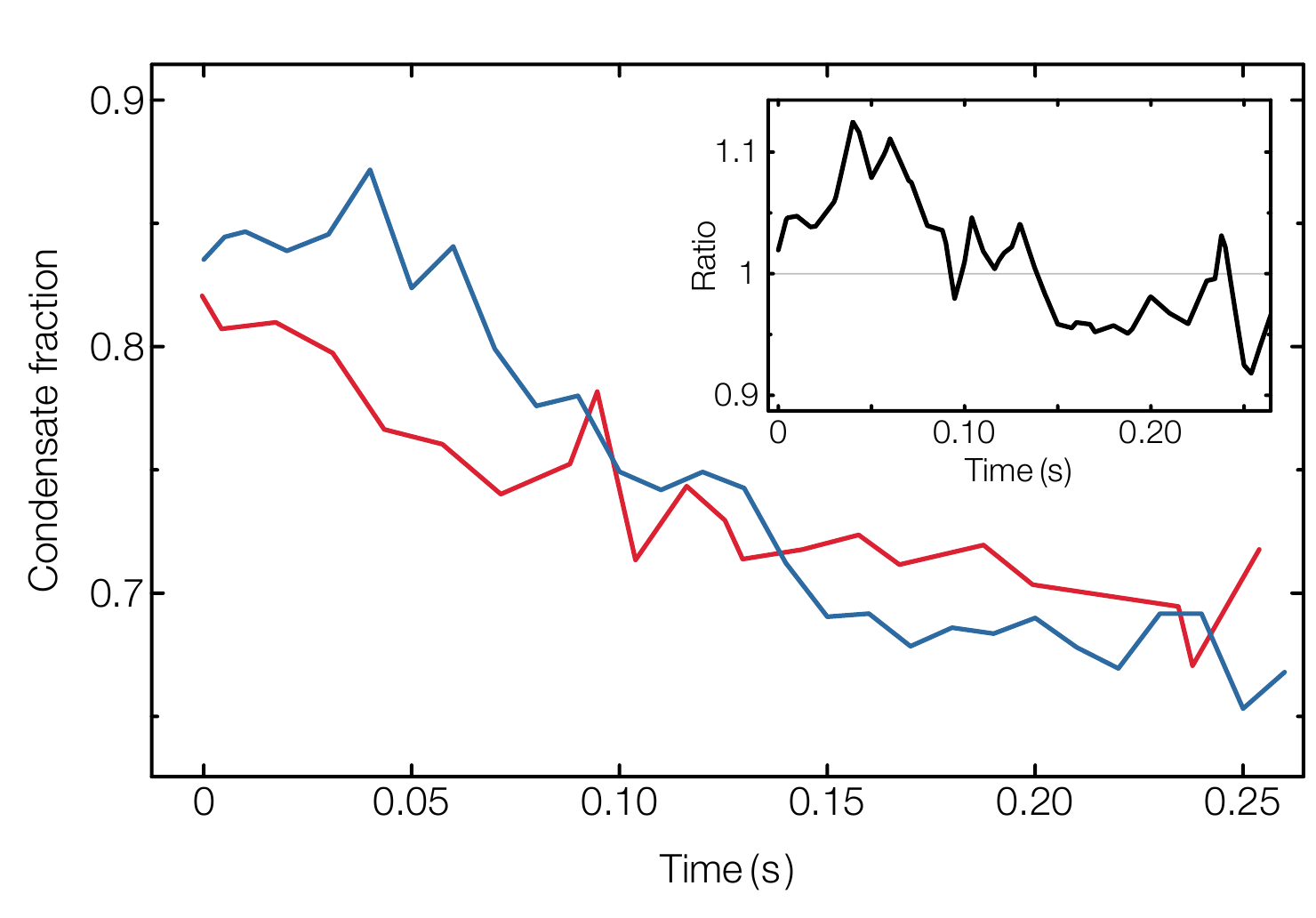}
\caption{Evolution of the condensed fraction for a molecular BEC in the absence of the barrier (red), and in the presence of a barrier of height $V_0/\mu \simeq 1.3$ (blue) and with $z_0 \simeq 0.4$ ($\Delta\mu_0/\mu \simeq 0.32$). For such values of $z_0$ and $V_0$, the evolution of the population imbalance $z(t)$ consists of a purely dissipative decay to zero. The inset displays the evolution of the ratio between the two measured condensed fractions. The condensed fraction appears to be limited only by the molecular BEC lifetime, with no markedly different trend detected in the presence of the barrier potential.}
\label{CondFrac}
\end{figure}
In our simulations (see Section \ref{ETFMmod}), we observe the instability of vortices towards the decay into short-wavelength density modulations. However, sound-like waves are not allowed to dissipate into thermal energy within the $T=0$ extended Thomas-Fermi model. Experimentally, we monitor the condensate fraction during the dissipative dynamics to reveal any possible significant heating of the system. The condensate fraction is extracted by fitting the recorded density profile of the expanding gas with a bi-modal distribution.
As shown in Fig.~\ref{CondFrac}, we do not observe any significant change in the evolution of the condensate fraction with respect to that recorded in the absence of the potential barrier (and thus in the absence of vortex nucleation). We conclude that no significant observable heating occurs during the dynamics.

\section{\mbox{Current-bias relation across the junction}}
We characterize the dependence of the dissipative current $I(t)$ as a function of the chemical potential difference across the barrier $\Delta \mu (t)$. %
By performing a numerical derivative of the time-evolving population imbalance, we obtain the instantaneous current $I(t)$ as a function of the instantaneous bias potential $\Delta \mu(t)$, which corresponds to the current-voltage relation of the equivalent circuit. In Fig.~\ref{figSM1}(a), the population imbalance $z(t)$ is shown for a gas at unitarity, with $z_0 \simeq 0.45$ and $V_0/\mu\simeq1.35$. As shown in Fig.~\ref{figSM1}(b), $I(t)$ exhibits a linear dependence on $\Delta \mu$, associated with an \textit{ohmic} dissipative current. The observed linear behaviour of $I(\Delta \mu)$ rules out any significant non-linear dissipation effects, differently for instance from what reported in Ref.~\cite{Brantut_2015}. There, a non-linear current-bias relation was observed for unitary superfluids connected via a quantum point-contact and attributed to multiple Andreev reflections.

\section{Resistively-shunted junction circuit}
\begin{figure}[t!]
\center
\vspace*{-8pt}
\includegraphics[width= \columnwidth]{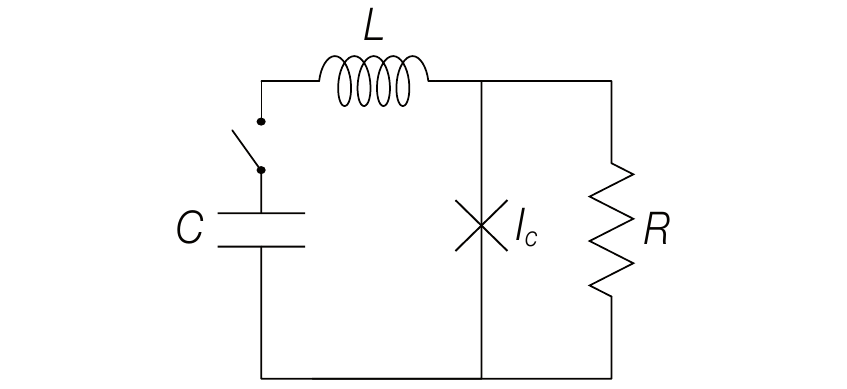}
\caption{Diagram of the RSJ-like circuit model described by Eqs.~\eqref{Kirchoff}-\eqref{JosAnd}. The Josephson junction allows a dissipationless current to flow up to a value of $I_c$, while an additional dissipative current is allowed to flow through the resistor $R$.}
\label{figRSJ}
\end{figure}

\begin{figure*}
\center
\includegraphics[width= 180mm]{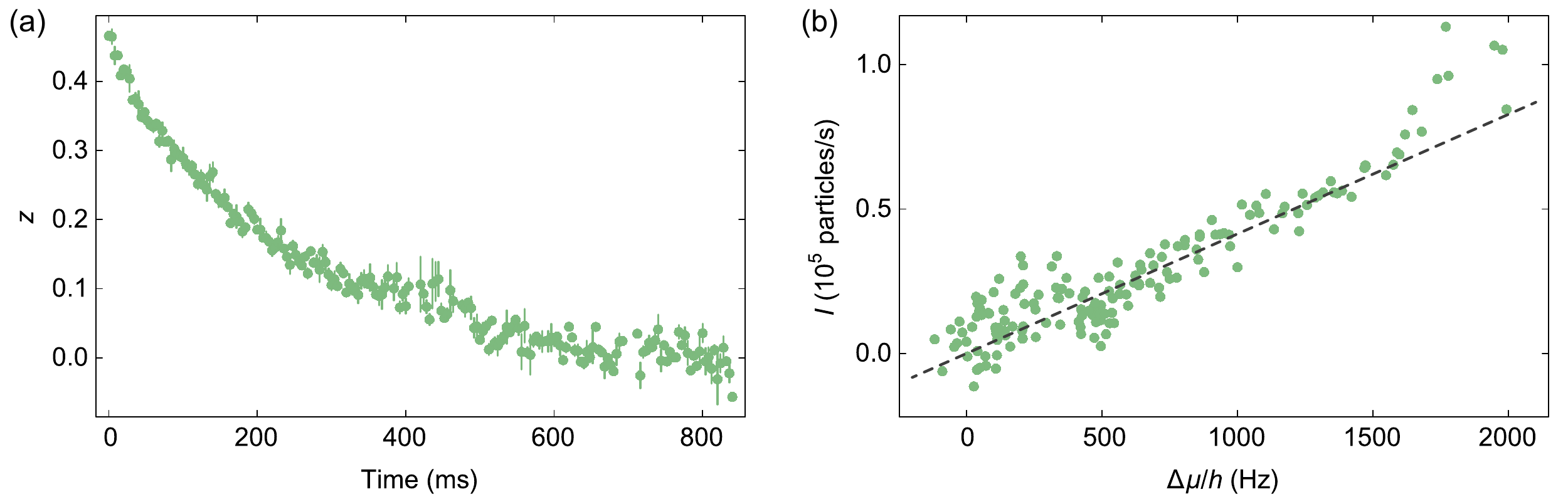}
\caption{Current-bias relation for a superfluid in the strongly dissipative regime. (a) Evolution of the population imbalance $z(t)$ for a unitary Fermi gas with $z_0 \simeq 0.45$ and $V_0/\mu\simeq1.35$. (b) Particle current $I$ as a function of $\Delta\mu/h$ obtained by numerical derivation of the experimental data shown in (a). The dashed black line is a linear fit to the data.}
\label{figSM1}
\end{figure*}

In order to characterize the transport properties of our junction, we model its dynamics using a RSJ-like circuit made of three parallel elements: a Josephson weak link with a current-phase relation \mbox{$I_J=-I_c\sin(\varphi)$}, a shunt resistance $R$ and a series $LC$ (see Ref.~\cite{Eckel2016} and Fig.~\ref{figRSJ}). The capacitance channel is associated with the potential energy stored in the junction, with $C={1 \over 2} {\partial N \over \partial\mu}$, i.e.~the gas compressibility \cite{Lee2013}, that it is evaluated using the superfluid equation of state \cite{MITEoS}. The inductance $L$ represents the kinetic energy of the atoms trapped in the harmonic  potential, and it is obtained experimentally by measuring $\omega_x=1/\sqrt{LC}$. The circuit is described by two coupled differential equations for $k(t)$ and $\varphi(t)$ \cite{Eckel2016}:
\begin{align}
L\ddot{k} +  R (\dot{k} + I_c \sin \varphi) + k/C &= 0\,,\label{Kirchoff}\\ 
\hbar \dot{\varphi} + R (\dot{k} + I_c \sin \varphi) &= 0 \label{JosAnd}.
\end{align}
Eq.~\eqref{Kirchoff} represents the circuit Kirchhoff's law, while Eq.~\eqref{JosAnd} is the generalized Josephson-Anderson relation. 
By numerically solving Eqs.~\eqref{Kirchoff}-\eqref{JosAnd}, we can obtain $z(t)$ and $\varphi(t)$. We fit the measured evolution of $z$ with the calculated one, leaving $R$ and $I_c$ as fitting parameters; in this way, we also obtain the corresponding evolution of $\phi$ for the best fit parameters (see Fig.~1 in the main text). 

\begin{figure*}
\center
\vspace*{5mm}
\includegraphics[width= 145mm]{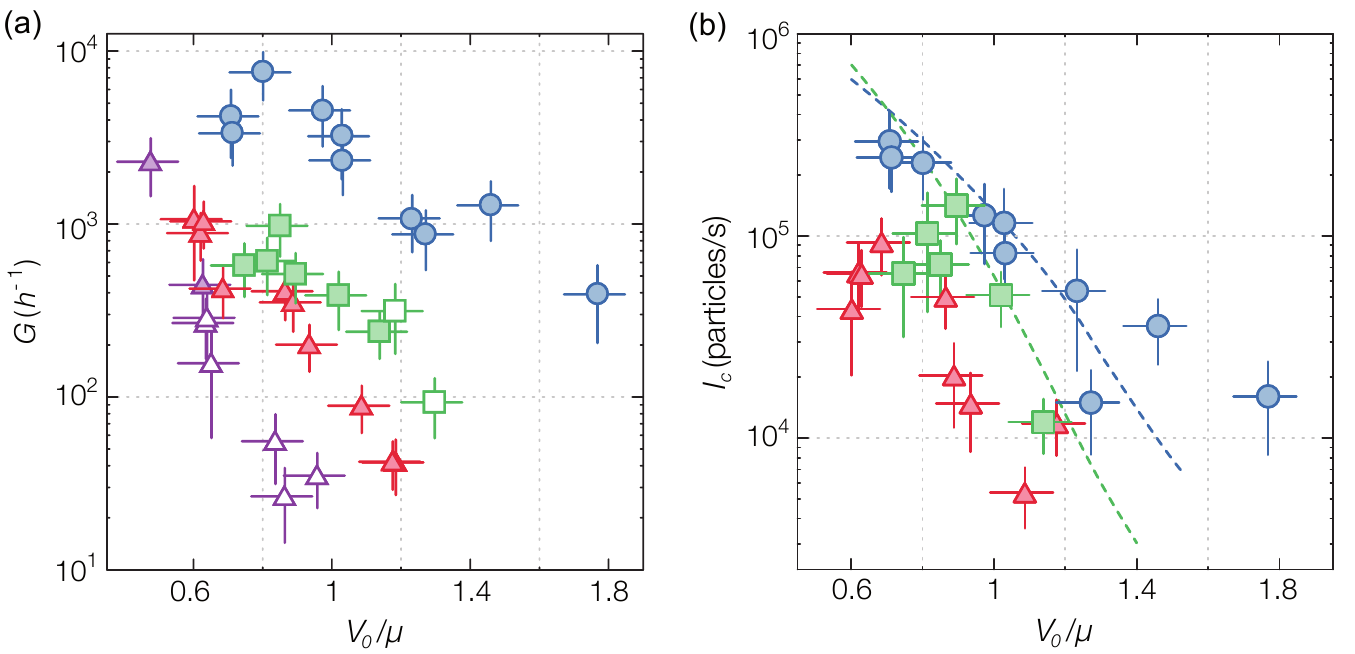}
\caption{(a) Conductance $G$ as a function of the normalized barrier height $V_0/\mu$ for molecular BECs at $1/(k_F a) = 4.6$ (blue circles), unitary Fermi gases (green squares), BCS superfluids at $1/(k_F a) = -0.6$ (red triangles), and an attractive Fermi gas at $1/(k_F a) = -1$ (purple triangles). For filled (empty) symbols, $G$ is obtained through a RSJ-like (RC) circuit model. (b) Critical current $I_c$ as a function of the normalized barrier height $V_0/\mu$ (same symbols as in (a)). %
The dashed curves denote the calculated critical current upper bounds $I_{c0}$ (see Section~\ref{critcurrtheory}), shown for BEC (blue) and unitary (green) superfluids. In both panels, the horizontal error bars are given by the typical $20\%$ uncertainty on the total atom number, while the vertical ones combine this with standard fitting errors.
}
\label{figIc}
\vspace*{2mm}
\end{figure*}

\section{Conductance and critical current as a function of barrier height}
By fitting the measured relative imbalance evolution $z(t)$ with the numerical solution of the RSJ-like circuit model in Eqs.~\eqref{Kirchoff}-\eqref{JosAnd} (see Fig.~\ref{figRSJ}), we extract the conductance $G$ and critical current $I_c$ of the junction. Figure~\ref{figIc} displays the extracted values of $G$ and $I_c$ as a function of the barrier strength $V_0/\mu$ in the different regimes of superfluidity, for $0.05 < \Delta\mu_0 < 0.2$. 
In the main text, we present instead the dependence of $G$ and $I_c$ upon the central density $n_0$, which is obtained by solving Eq.~\eqref{ETFM_eq} at a given $V_0/\mu$ (see Section~\ref{ETFMmod}).  
At a fixed $V_0/\mu$, $G$ decreases from the BEC to the BCS side of the crossover. %
On the other hand, at equal central density $n_0$ the different superfluids exhibit approximately matching conductances over all the explored range of barrier strengths (see Fig.~3 in the main text). In Fig.~\ref{figIc}(a), the conductance of an attractive Fermi gas at $1/(k_F a) = -1$ is also displayed for comparison. The superfluid fraction of our sample at $1/(k_F a) = -1$ is small, since the gas temperature $T \simeq 0.1\,T_F$ corresponds to $T/T_c \sim 1$. Similarly to what observed for a non-interacting Fermi gas \cite{valtolina}, no Josephson oscillations are detected for barrier strengths larger than $V_0/\mu \simeq 0.6$ (empty symbols), corresponding roughly to the mean energy per particle at $1/(k_F a) = -1$. The extracted $G$ lies well below that of BCS superfluids at $1/(k_F a) = -0.6$ despite the very similar chemical potential $\mu$, with a 10-fold reduction at $V_0/\mu \simeq 1$. These observations point to a different microscopic mechanism dominating the conduction in non-superfluid samples, most likely associated with single-particle tunnelling. Figure~\ref{figIc}(b) shows that, for all superfluids, $I_c$ decreases upon increasing $V_0$, as expected by the exponential reduction of the tunnelling strength (shaded lines in the figure denote exponential fits to the data). The calculated upper bounds $I_{c0}$ for the critical current in the weak barrier regime (see Section~\ref{critcurrtheory}) are also plotted (dashed lines). 

\section{Theoretical methods:\\the extended Thomas-Fermi model}\label{ETFMmod}
We use the extended Thomas-Fermi model (ETFM) \cite{Forbes_2014} for determining the bulk properties of the gas and for theoretically investigating the onset of dissipation at $T=0$. This model is an extension of the Gross-Pitaevskii equation (GPE) for atom pairs, where the local chemical potential is parametrized as a function of the scattering length $a$, according to the equation of state in the BEC-BCS crossover \cite{Gan11}. In this framework, the condensate wave function $\psi(\textbf{r},t)$, normalized to the total number of condensate pairs $N$, obeys the following equation:

\begin{equation}
i \hbar \partial_{t}\psi(\textbf{r},t)=\left(-\frac{\hbar^2}{2M} 
\nabla^2+V(\textbf{r})+f(\vert \psi(\textbf{r},t) \vert^2,a)\right)\psi(\textbf{r},t)\,,
\label{ETFM_eq}
\vspace*{5pt}
\end{equation}

\noindent where $M=2m$ is the mass of an atomic pair, $V(\textbf{r})$ is the trapping potential and $f(\vert \psi(\textbf{r},t) \vert^2, a)$ is the local chemical potential. On the BEC side of the Feshbach resonance, for small repulsive interactions, it holds that $f\left(\vert \psi(\textbf{r},t) \vert^2,a\right)\rightarrow g \left|\psi(\textbf{r},t)\right|^{2}$, where $g=4\pi\hbar^{2}a_{M}/M$, and $a_{M}=0.6\,a$ is the inter-molecular scattering length. In this limit, the ETFM coincides with the GPE for weakly interacting bosonic particles. On the other hand, at unitarity $a$ diverges and it disappears from the equation of state, which depends only on the Bertsch parameter $\xi$. Here, one has that $f(\vert \psi(\textbf{r},t) \vert^2)=\alpha(\left|\psi(\textbf{r},t)\right|^{2})^{2/3}$, where $\alpha=2\xi\,\hbar^{2}/M\,(6\pi^{2})^{2/3}$, and $\xi$ is set to the experimentally determined value $\xi = 0.37$ \cite{Zwierlein_2012,Zurn2013}.

\subsection{Equilibrium properties of the superfluid}

\begin{figure}[t!]
\center
\vspace*{-7pt}
\includegraphics[width= \columnwidth]{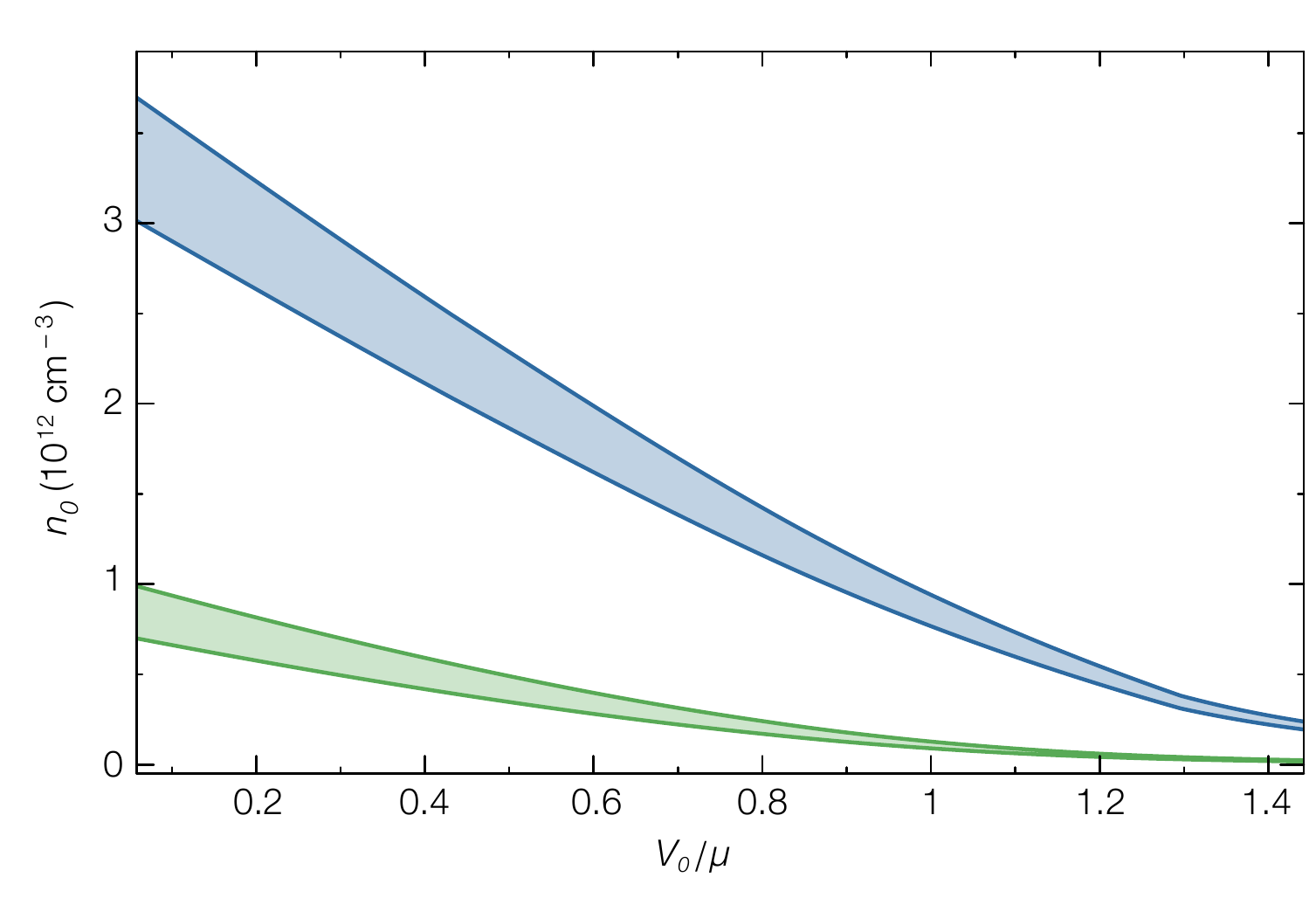}
\caption{Central density $n_0$ as a function of $V_0/\mu$, calculated through the ETFM in Eq.~\eqref{ETFM_eq}. The curves shown refer to the molecular BEC at 1/($k_Fa)=4.6$ (blue) and unitary gas (green). The shaded regions delimit the typical intervals of atom numbers used in the experiment: $N = 0.6 - 1 \times 10^5$ pairs for BEC superfluids, and $N = 1 - 2 \times 10^5$ pairs for unitary superfluids.}
\label{calibn0}
\end{figure}
\begin{figure}[t!]
\center
\vspace*{0pt}
\includegraphics[width= \columnwidth]{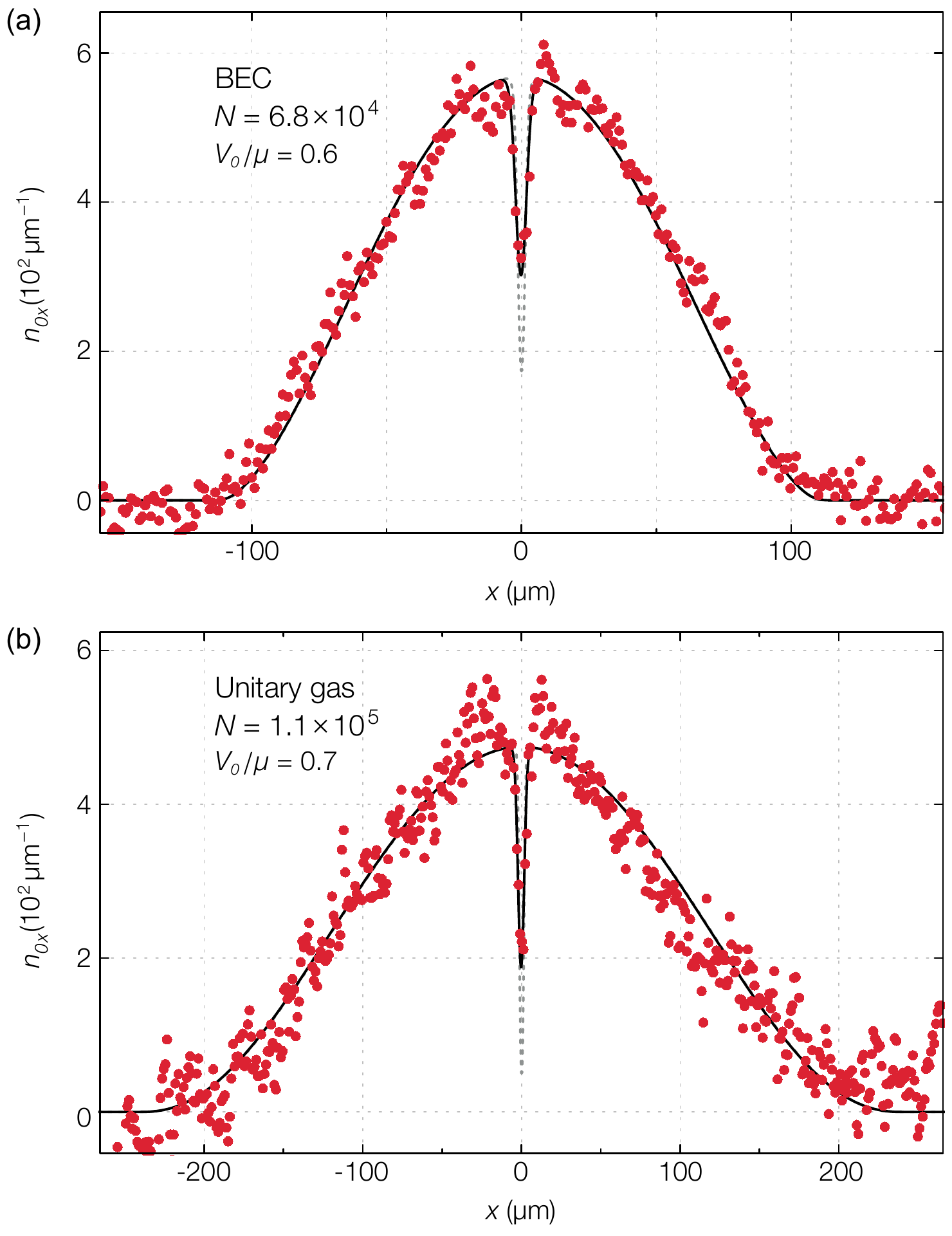}
\caption{Radially integrated density $n_{0x}$ as a function of the axial position $x$: (a) molecular BEC at 1/($k_Fa)=4.6$ with $V_0/\mu \simeq 0.6$ and $N \simeq 6.8 \times 10^4$, and (b) unitary Fermi gas with $V_0/\mu \simeq 0.7$ and $N \simeq 1.1 \times 10^5$. The experimental data points are obtained by integrating a single in-situ absorption image with $z_0\simeq0$ along the $y$ direction. The theoretical $n_{0x}(x)$ is obtained by numerically solving the ETFM in Eq.~\eqref{ETFM_eq} without any free parameters (dashed gray line). To include the effect of the imaging resolution, the calculated density profile is convoluted with the imaging PSF (solid black line).}
\label{density-profiles}
\end{figure}

Since the ETFM does not include any fermionic degree of freedom, it cannot correctly describe the system dynamics throughout the whole BEC-BCS crossover. Notwithstanding this limitation, it still provides a useful platform for evaluating the static properties of the superfluid \cite{Forbes_2014}. As already done in Ref.~\cite{valtolina}, we follow this approach to calculate the gas compressibility for different values of $1/\left(k_{F}a\right)$. Additionally, on the BEC side of the resonance and at unitarity, where the condensed fraction is still large, the ETFM yields a good estimation of the ground-state wave function $\psi_{0}(\textbf{r})$, which is obtained by solving Eq.~\eqref{ETFM_eq} with the initial condition $z_0=0$. This allows us to estimate the equilibrium pair density $n_0(\textbf{r})=\left|\psi_{0}(\textbf{r})\right|^{2}$ in the presence of the trapping potential of Eq.~\eqref{trapping_potential}.
Since the barrier parameters and the total pair number $N$ are obtained experimentally, %
$n_0(\textbf{r})$ can be derived without any free parameters. For the molecular BEC at $1/(k_F a) = 4.6$ and the unitary Fermi gas, we calculate the density in the center of the barrier potential, i.e.~ $n_0(\textbf{0})$ with $\textbf{0} = (0,0,0)$, for different values of $V_0/\mu$ (see Fig.~\ref{calibn0}). In this way, we can express the conductance $G$, which is extracted by fitting the experimental data with the circuit model for different values of the barrier strength $V_0/\mu$ (see Fig.~\ref{figIc}), as a function the central density $n_0 \equiv n_0(\textbf{0})$ (see Fig.~3 in the main text). This method is more accurate than experimentally extracting $n_0$ from the recorded in-situ absorption images and inverse Abel transform: the aperture of our imaging system and the effect of atom diffusion during the typical $5\,\mu$s-long imaging pulse ultimately limit the imaging resolution to about $1.5\,\mu$m, compromising the accuracy of such measurement for barrier heights $V_0 \sim \mu$.  
To directly validate the ETFM calculation of the central density at equilibrium, we compare the experimental in-situ density profiles to the calculated radially integrated density profiles $n_{0x}(x) = \int dy\,dz\,n_0(\textbf{r})$, taking into account the effect of the finite imaging resolution. To this purpose, we perform a convolution between the calculated density profiles and a Gaussian function with a 1/$e^2$ radius of 3\,$\mu$m, roughly corresponding to the imaging point-spread function HWHM of 1.5\,$\mu$m. The result of such a comparison is displayed in Fig.~\ref{density-profiles}, showing good agreement between the data and the calculated profiles, both for BEC and unitary superfluids. For the BCS superfluid at $1/(k_F a) = - 0.6$, we roughly estimate the central density by rescaling the values previously calculated for the unitary gas, using a constant scaling factor given by the ratio between the maximum density of the BCS gas and that of the unitary gas confined in a purely harmonic potential.

\subsection{Theoretical estimation of the critical current}\label{critcurrtheory}

We provide an estimate for the critical current of both the BEC and unitary gas by setting, as an upper bound for the superfluid critical velocity, the average local speed of sound $c$ in the plane of the junction. It has been numerically shown that a bosonic superfluid flowing through a barrier becomes unstable above this kinematic threshold: once the superfluid velocity exceeds $c$, vortex excitations are nucleated inside the barrier at the edge of the superfluid \cite{piazza}. For bosonic and unitary superfluids confined in an elongated trap, as in our case, $c$ is given by $c_0/\sqrt{2}$ \cite{Zaremba} and $c_0\sqrt{3/5}$ \cite{Tosi}, respectively. Here, $c_0$ denotes the speed of sound in a homogeneous superfluid, and it is given by $c_0= \sqrt{(\gamma \beta/M)(n_0(\textbf{r}_{\textbf{}0}))^{\gamma}}$ with $\beta=g$ and $\gamma=1$ for a Bose gas, and $\beta=2\xi\,\hbar^{2}/M\,(6\pi^{2})^{2/3}$ and $\gamma=2/3$ for a unitary Fermi gas. An upper bound for the critical current $I_c$ is then derived using the hydrodynamic relation $I_{c0}=n_{0x}\,c$, where $n_{0x} \equiv n_{0x}(0)$ is the radially integrated pair density at the center of the barrier potential. Therefore, $I_{c0}$ depends on the nature of the superfluid considered, on the central density $n_0$, and on the effective area of the junction, which is associated to the radial size of the superfluid. Within this calculation, the superfluid density is assumed to be equal to the condensate one, completely neglecting quantum depletion. 
Such crude approximation yields only an upper bound for $I_c$. Nonetheless, for BEC and unitary superfluids, we find that the calculated $I_{c0}$ values to lie reasonably close to our data points, for which $I_c$ is extracted using the circuit model (see Fig.~3(b) in the main text and Fig.~\ref{figIc}(b)). We thus conclude that the larger values of $I_c$ at fixed pair density observed at unitarity is due, in addition to the increase of the area of the junction, to the rise of $c$. This is expected since the speed of sound in the fluid is maximum for resonant interactions. 
On the other hand, moving to the BCS side of the crossover, even if both the speed of sound and the radii of the cloud further increase, we do not observe a detectable increase of the critical current, but rather a reduction.  
This behaviour could be attributed to the decrease of the critical superfluid velocity, which in the BCS limit is bounded by the pair-breaking velocity \cite{Spuntarelli}. Deep in the tunnelling regime, where the hydrodynamic relation for the critical current is no longer reliable, the reduction of $I_c$ moving from unitarity to the BCS limit is expected as a consequence of increasing condensate depletion, as shown in Ref.~\cite{valtolina}. These complementary pictures, which are correct only in the two limiting regimes of transport, are consistent within one another, since both the critical velocity for superfluid instability and the condensate fraction are ultimately determined by the fermionic-pairing gap.

\subsection{Simulation of the junction dynamics}
\begin{figure}[t!]
\center
\includegraphics[width=\columnwidth]{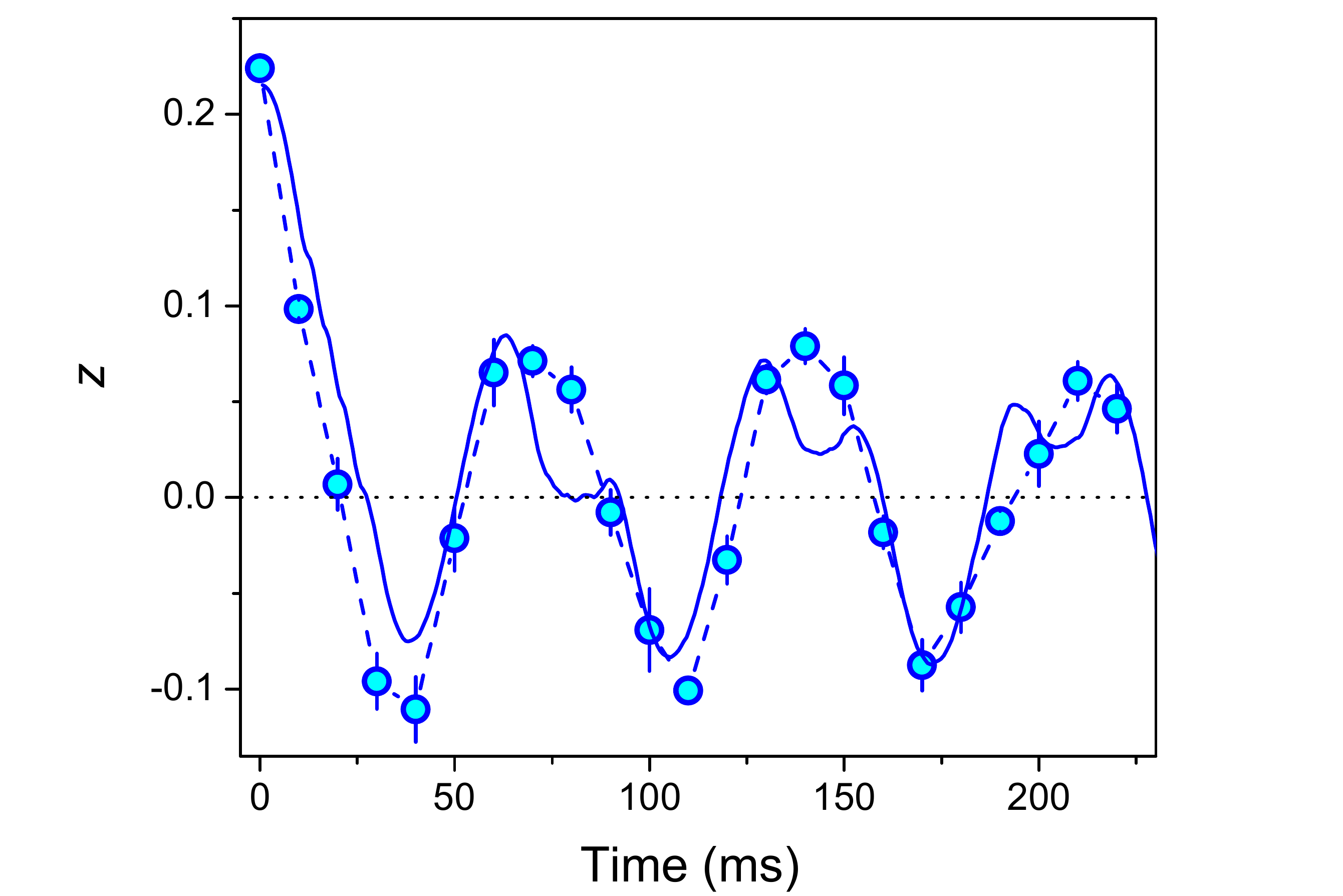}
\caption{Relative population imbalance $z$ (blue circles) measured as a function of time for a molecular BEC at $1/(k_F a) = 4.6$ and $V_0/\mu \simeq 0.7$. The solid line is the corresponding ETFM predictions for a BEC. In the numerical simulation, the particle number is set to $N=6\times10^4$, in agreement with the experimental conditions.}
\label{GPE_dynamics}
\end{figure}

In Ref.~\cite{valtolina}, the $T=0$ ETFM was already shown to correctly predict the Josephson plasma frequency for BEC and unitary superfluids at small initial population imbalance $z_0$. Here, we use the ETFM for simulating the junction dynamics once the initial population imbalance $z_0$ is increased above a critical threshold $z_c$ for the emergence of dissipation. For the molecular BEC, we find that the model predicts properly the value of $z_c$ and qualitatively reproduces the evolution of $z$ for $V_0<\mu$. In Fig.~\ref{GPE_dynamics}, we compare the ETFM simulation with the experimental data reported in Fig.~1(c) of the main text. In the simulation, the initial population imbalance is created following the procedure detailed in Ref.~\cite{valtolina}. In both the simulated and experimental evolution, we observe an initial decay of $z$ followed by Josephson plasma oscillations. We point out that the theory does not contain any free parameter or additional \textit{ad hoc} dissipative terms, and the reasonable agreement between data and simulations demonstrates therefore that the resistive particle flow does not arise from thermal excitations. 

\begin{figure*}
\center
\includegraphics[width= 179mm]{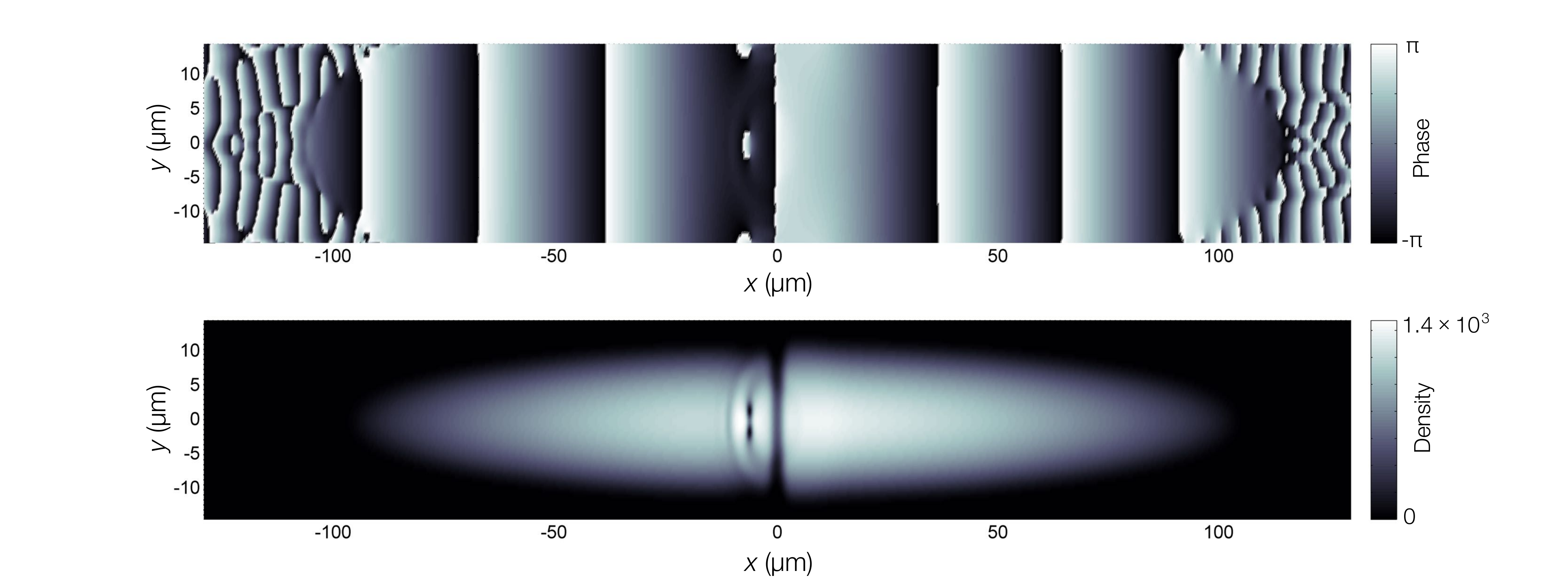}
\caption{Snapshots of the simulated condensate phase (top) and density (bottom), in the $xy$~plane, for a molecular BEC at $1/(k_F a) = 4.6$ after an evolution time of 13.6~ms, with $z_0=0.2$, $V_{0}/\mu=0.7$ and $N=6\times10^4$. The pair density is expressed in dimensionless form using $a_x^{3}$ as volume unit, with $a_x=\sqrt{\hbar/M\omega_{x}}$ being the harmonic oscillator length. In both density and phase contours, a vortex ring is visible in the left reservoir.}
\label{BEC}
\end{figure*}
\begin{figure*}
\center
\includegraphics[width=179mm]{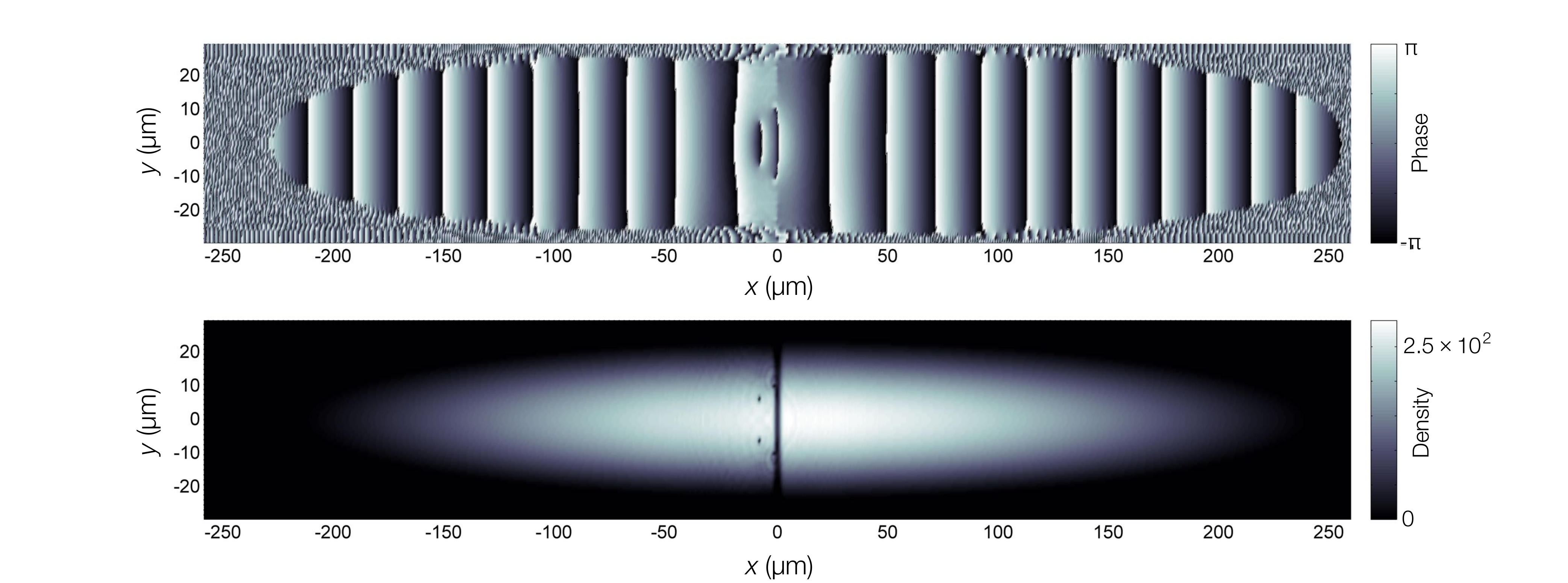}
\caption{Snapshots of the simulated condensate phase (top) and density (bottom), in the $xy$~plane, for a unitary Fermi gas after an evolution time of 10.2~ms, with $z_0=0.2$, $V_{0}/\mu=0.5$ and $N=8\times10^4$. The pair density is expressed in dimensionless form using $a_x^{3}$ as volume unit, with $a_x=\sqrt{\hbar/M\omega_{x}}$ being the harmonic oscillator length. Two vortex rings, one on the left side of the barrier and  another just detaching from the barrier, are observed in both density and phase contours.}
\label{UFG}
\vspace*{3pt}
\end{figure*}
The $T=0$ model has however an intrinsic limitation in comprehensively describing dissipative processes, due to impossibility of exchanging or transferring energy to a thermal bath. As a matter of fact, the starting excitation energy in the simulations is completely converted into density modulations, as shown in Fig.~\ref{GPE_dynamics}, where in addition to the Josephson plasma mode, we observe the presence of other dynamical modes, which are not resolved in the experiment. Therefore, even if the superfluid instability is not caused by thermal effects, a complete theoretical description should include the dynamical coupling of the condensate to the thermal cloud. As we move from the BEC to the unitary limit, the predictions of the ETFM start to deviate from the experimental results. In particular, even though the general trend of an initial decay followed by an oscillating behavior is confirmed, the model does not correctly estimate the value of $z_c$. This is not surprising because the quantum depletion of the condensate and finite-temperature effects could start playing a role that can no more be completely neglected.
\smallskip
Despite these shortcomings, it is enlightening to investigate the microscopic mechanism underlying the particle current decay observed in the ETFM simulations, both in BEC and unitary regime. Previous simulations of three-dimensional weakly linked bosonic superfluids already showed that phase slips arise from vortex rings nucleated within the barrier at the cloud edge and oriented perpendicular to the flow \cite{piazza, PiazzaJOPB, Abad}. The combination of low density and inhomogeneity in the barrier region causes the vortex core to shrink radially in size, crossing the junction region \cite{piazza}. In Fig.~\ref{BEC} and Fig.~\ref{UFG}, we show some typical results of ETFM simulations of vortex dynamics. The pair density $n(x,y)=\left|\psi(x,y,0)\right|^{2}$ and the phase of the order parameter $\Phi(x,y,0)$ are displayed for a molecular BEC and a unitary gas after an evolution time of 13.6\,ms and 10.2\,ms, respectively. In both cases $z_0$ is set to $0.2$ and $V_{0}/\mu$ to $0.7$ for the BEC and to $0.5$ for the unitary gas. 
The simulations reveal the presence of vortex rings appearing as vortex-antivortex pairs in the horizontal plane. 

\noindent The topological defects are nucleated at the edge of the superfluid in the barrier region, and enter into the bulk by shrinking in size, giving rise to phase slips across the junction. The phase-slippage rate, which is proportional to $\Delta\mu$, is consistent with the number of observed vortices for the BEC and the unitary Fermi gas. For the BEC (see Fig.~\ref{BEC}), only one vortex ring moves along the direction of the flow, corresponding to the first phase slip. In the unitary gas (see Fig.~\ref{UFG}), while the first vortex ring propagates into the bulk, a second one starts to detach from the barrier. The increase of vortex population moving from the BEC limit to unitarity is a consequence of the increase of $\Delta\mu$ at fixed $z_0$. 

Our numerical results show that, for sufficiently high barriers, vortex rings are unstable: after entering into the bulk, they shrink and disappear. We have also investigated the effect of adiabatically removing the barrier, following the experimental procedure for vortex detection (see Fig.~\ref{ramps}). We find that the barrier removal after nucleation stabilizes the vortices. In BEC and unitary superfluids, the number of vortices observed in the experiment completely decays within $\sim50$\,ms, with essentially no vortices detected during the Josephson plasma oscillations. On the other hand, we observe a sporadic survival of vortices in BCS superfluids at long evolution times (see Fig.~2(f) in the main text). While we lack a complete understanding of this feature, this may result from the combination of the higher nucleation rate and the different effective mass and core-size of vortices in the crossover region \cite{valtolina, MIT_solitons}. A detailed study of vortex dynamics and decay in BEC superfluids, with and without barrier, is presently ongoing \cite{Proukakis}.
We point out that, for both BEC and unitary gases, the agreement between theory and experiment is lost for barrier heights $V_0 > \mu$. In this range, the values of the conductance extracted from the simulations are finite but significantly lower than the experimentally measured ones. Furthermore, we find that in these conditions the vortex rings do not escape the barrier region and annihilate inside it. The only excitations propagating into the superfluid bulk are sound-like waves, leading to a very low conductance. The failure of the $T=0$ ETFM approach in the tunnelling regime suggests that both thermal and quantum fluctuations may significantly influence the system dynamics. %


\vspace*{-90pt}

\begin{thebibliography}{60}

\vspace*{35pt}

\bibitem{barone} A. Barone and G. Patern\`o, \textit{Physics and applications of the Josephson effect} (Wiley, New York, 1982).

\bibitem{Tinkham} M. Tinkham, \textit{Introduction to Superconductivity}, 2nd ed. (McGraw-Hill, New York, 1996).

\bibitem{Halperin} B. I. Halperin, G. Refael, and E. Demler, \textit{Int. J. Mod. Phys. B} \textbf{24}, 4039 (2010).

\bibitem{Var14} E. Varoquaux, \textit{Rev. Mod. Phys.} \textbf{87}, 803 (2015).

\bibitem{Feynman} R. P. Feynman, in \textit{Progress in Low Temperature Physics}, edited by C. J. Gorter (North-Holland, Amsterdam, 1955). 

\bibitem{And66} P. W. Anderson, \textit{Rev. Mod. Phys.} \textbf{38}, 298 (1966).

\bibitem{avenel1985} O. Avenel and E. Varoquaux, \textit{Phys. Rev. Lett.} \textbf{55}, 2704 (1985).

\bibitem{Langer66} J. S. Langer and V. Ambegaokar,  \textit{Phys. Rev.} \textbf{164}, 498 (1967).

\bibitem{Bezryadin} A. Bezryadin, C.N. Lau, and M. Tinkham, \textit{Nature} \textbf{404}, 971 (2000).

\bibitem{Kamenev2014} Y. Chen, Y.-H. Lin, S. D. Snyder, A. M. Goldman, and A. Kamenev, \textit{Nature Phys.} \textbf{10}, 567 (2014).

\bibitem{SQUID} J. Clarke and A. I. Braginski, \textit{The SQUID Handbook} (Wiley-VCH, Weinheim, 2004).

\bibitem{Packard2012} Y. Sato and R.E. Packard, \textit{Rep. Prog. Phys.}, \textbf{75}, 016401 (2012). 

\bibitem{Ventra} C.-C. Chien, S. Peotta and M. Di Ventra, \textit{Nature Phys.} \textbf{11}, 998 (2015).

\bibitem{Lev07} S. Levy, E. Lahoud, I. Shomroni, and J. Steinhauer, \textit{Nature} \textbf{449}, 579 (2007).

\bibitem{DeMarco2008} D. McKay, M. White, M. Pasienski, and B. DeMarco, \textit{Nature} \textbf{453}, 76 (2008).

\bibitem{Hadzibabic_2012} S. Moulder, S. Beattie, R. P. Smith, N. Tammuz, and Z. Hadzibabic, \textit{Phys. Rev. A} \textbf{86}, 013629 (2012). 

\bibitem{Wrigh_2013} K. C. Wright, R. B. Blakestad, C. J. Lobb, W. D. Phillips, and G. K. Campbell, \textit{Phys. Rev. Lett.} \textbf{110}, 025302 (2013). 

\bibitem{Jen2014} F. Jendrzejewski, S. Eckel, N. Murray, C. Lanier, M. Edwards, C. J. Lobb, G. K. Campbell, \textit{Phys. Rev. Lett.} \textbf{113}, 045305 (2014).

\bibitem{Tanzi2016} L. Tanzi, S. Scaffidi Abbate, F. Cataldini, L. Gori, E. Lucioni, M. Inguscio, G. Modugno, and C. D'Errico, \textit{Sci. Rep.} \textbf{6}, 25965 (2016).

\bibitem{Eckel2016} S. Eckel, J. G. Lee, F. Jendrzejewski, C. J. Lobb, G. K. Campbell, and  W. T. Hill, III, \textit{Phys. Rev. A} \textbf{93}, 063619 (2016).

\bibitem{Miller2007} D. E. Miller, J. K. Chin, C. A. Stan, Y. Liu, W. Setiawan, C. Sanner, and W. Ketterle, \textit{Phys. Rev. Lett.} \textbf{99}, 070402 (2007).

\bibitem{Watanabe2009} G. Watanabe, F. Dalfovo, F. Piazza, L. P. Pitaevskii, and S. Stringari, \textit{Phys. Rev. A} \textbf{80}, 053602 (2009).

\bibitem{Weimer2015} W. Weimer, K. Morgener, V. P. Singh, J. Siegl, K. Hueck, N. Luick, L. Mathey, and H. Moritz, \textit{Phys. Rev. Lett.} \textbf{114}, 095301 (2015).

\bibitem{Delehaye2015} M. Delehaye, S. Laurent, I. Ferrier-Barbut, S.Jin, F. Chevy, and C. Salomon, \textit{Phys. Rev. Lett.} \textbf{115}, 265303 (2015) .

\bibitem{Castin2015} Y. Castin, I. Ferrier-Barbut, and C. Salomon, \textit{C. R. Phys.} \textbf{16}, 241 (2015).

\bibitem{Sta12} D. Stadler, S. Krinner, J. Meineke, J.P. Brantut, and T. Esslinger, \textit{Nature} \textbf{491}, 736 (2012).

\bibitem{husmann2015} D. Husmann, S. Uchino, S. Krinner, M. Lebrat, T. Giamarchi, T. Esslinger, and J. P. Brantut, \textit{Science} \textbf{350}, 1498 (2015).

\bibitem{valtolina} G. Valtolina, A. Burchianti, A. Amico, E. Neri, K. Xhani, J. A. Seman, A. Trombettoni, A. Smerzi, M. Zaccanti, M. Inguscio, and G. Roati, \textit{Science} \textbf{350}, 1505 (2015).

\bibitem{Bur14} A. Burchianti, G. Valtolina, J. A. Seman, E. Pace, M. De Pas, M. Inguscio, M. Zaccanti, and G. Roati, \textit{Phys. Rev. A} \textbf{90}, 043408 (2014).

\bibitem{SM} See Supplemental Material for details on the experimental procedures, and the theoretical methods and simulations, which includes Refs.~[31--36].

\bibitem{MIT_solitons} M. J. H. Ku, W. Ji, B. Mukherjee, E. Guardado-Sanchez, L. W. Cheuk, T. Yefsah, and M. W. Zwierlein, \textit{Phys. Rev. Lett.} \textbf{113}, 065301 (2014).
\bibitem{Lee2013}  J. G. Lee, B. J. McIlvain, C. J. Lobb, and W. T. Hill, III, \textit{Sci. Rep.} \textbf{3}, 1034 (2013).
\bibitem{MITEoS} M. J. H. Ku, A. T. Sommer, L. W. Cheuk, M. W. Zwierlein, \textit{Science} \textbf{335}, 563 (2012).
\bibitem{Gan11} S. Gandolfi, K. E. Schmidt, and J. Carlson, \textit{Phys. Rev. A} \textbf{83}, 041601(R) (2011).
\bibitem{Zaremba} E. Zaremba, \textit{Phys. Rev. A}  \textbf{57}, 518 (1998).
\bibitem{Tosi} P. Capuzzi, P. Vignolo, F. Federici, and M. P. Tosi, \textit{Phys. Rev. A} \textbf{73}, 021603(R) (2006).

\bibitem{zurn2013} G. Z\"urn, T. Lompe, A. N. Wenz, S. Jochim, P. S. Julienne, and J. M. Hutson, \textit{Phys. Rev. Lett.} \textbf{110}, 135301 (2013). 

\bibitem{Sme97} A. Smerzi, S. Fantoni, S. Giovanazzi, and S. R. Shenoy, \textit{Phys. Rev. Lett.} \textbf{79}, 4950 (1997).

\bibitem{Zapata} I. Zapata, F. Sols, A. J. Leggett, \textit{Phys. Rev. A} \textbf{57}, R28(R) (1998).

\bibitem{Meier2001} F. Meier and W. Zwerger, \textit{Phys. Rev. A} \textbf{64}, 033610 (2001).

\bibitem{Zou2014} P. Zou and F. Dalfovo, \textit{J. Low Temp. Phys.} \textbf{177}, 240 (2014).

\bibitem{Alb05} M. Albiez, R. Gati, J. F\"{o}lling, S. Hunsmann, M. Cristiani, and M. K. Oberthaler, \textit{Phys. Rev. Lett.} \textbf{95}, 010402 (2005).

\bibitem{Abbarchi2013} M. Abbarchi, A. Amo, V. G. Sala, D. D. Solnyshkov, H. Flayac, L. Ferrier, I. Sagnes, E. Galopin, A. Lematre, G. Malpuech, and J. Bloch, \textit{Nature Phys.} \textbf{9}, 275 (2013).

\bibitem{Fattori2016} G. Spagnolli, G. Semeghini, L. Masi, G. Ferioli, A. Trenkwalder, S. Coop, M. Landini, L. Pezze', G. Modugno, M. Inguscio, \textit{et al.}, \textit{Phys. Rev. Lett.} \textbf{118}, 230403 (2017).

\bibitem{Ruostekoski1998} J. Ruostekoski and D.F. Walls, \textit{Phys. Rev. A} \textbf{58}, R50 (1998).

\bibitem{Suthar} K. Suthar, A. Roy and D. Angom, \textit{J. Phys. B: At. Mol. Opt. Phys.} \textbf{47}, 135301 (2014).

\bibitem{proukakis} Further theoretical investigations of vortex dynamics at finite temperature are ongoing, in collaboration with Prof.~N.~Proukakis.

\bibitem{piazza} F. Piazza, L. A. Collins, and A. Smerzi, \textit{New J. Phys.}, \textbf{13}, 043008 (2011).

\bibitem{PiazzaJOPB} F. Piazza, L. A. Collins, and A. Smerzi, \textit{J. Phys. B: At. Mol. Opt. Phys.}, \textbf{46}, 095302 (2013).

\bibitem{Abad} M. Abad, M. Guilleumas, R. Mayol, F. Piazza, D. M. Jezek, and A. Smerzi, \textit{Eur. Phys. Lett.} \textbf{109}, 40005 (2015).

\bibitem{Forbes_2014} M. McNeil Forbes and R. Sharma, \textit{Phys. Rev. A} \textbf{90}, 043638 (2014).

\bibitem{MIT_solitons} M. J. H. Ku, B. Mukherjee, T. Yefsah, and M. W. Zwierlein, \textit{Phys. Rev. Lett.} \textbf{116}, 045304 (2016).

\bibitem{Stewart68} W. C. Stewart, \textit{Appl. Phys. Lett.} \textbf{12}, 277 (1968).

\bibitem{McCumber68} D. E. McCumber, \textit{J. Appl. Phys.} \textbf{39}, 3113 (1968).

\bibitem{Giaever} I. Giaever, \textit{Phys. Rev. Lett.} \textbf{5}, 464 (1960).

\bibitem{Spuntarelli} A. Spuntarelli, P. Pieri, and G. C. Strinati, \textit{Phys. Rev. Lett.} \textbf{99}, 040401 (2007).

\bibitem{Tsatsos} M. C. Tsatsos, P. E.S. Tavares, A. Cidrim, A. R. Fritsch, M. A. Caracanhas, F. E. A. dos Santos, C. F. Barenghi, V. S. Bagnato, \textit{Phys. Rep.} \textbf{622}, 1 (2016).

\bibitem{Bulgac} A. Bulgac, M. M. Forbes, and G. Wlazlowski, \textit{J. Phys. B} \textbf{50}, 014001 (2017).

\bibitem{trentoPRX} S. Serafini, L. Galantucci, E. Iseni, T. Bienaim\'e, R. N. Bisset, C. F. Barenghi, F. Dalfovo, G. Lamporesi, and G. Ferrari, \textit{Phys. Rev. X} \textbf{7}, 021031 (2017).

\bibitem{Zhai} B. Liu, H. Zhai, and S. Zhang, \textit{Phys. Rev. A} \textbf{90}, 051602(R) (2014).

\bibitem{caldeira} A. O. Caldeira and A. J. Leggett, \textit{Ann. Phys.} \textbf{149}, 374 (1983).

\bibitem{fisher} M. P. A. Fisher, \textit{Phys. Rev. Lett.} \textbf{57}, 885 (1986).

\end{thebibliography}

\begin{thebibliography}{30}
\bibitem{valtolina} G. Valtolina, A. Burchianti, A. Amico, E. Neri, K. Xhani, J. A. Seman, A. Trombettoni, A. Smerzi, M. Zaccanti, M. Inguscio, and G. Roati, \textit{Science} \textbf{350}, 1505 (2015).
\bibitem{Burchianti2014} A. Burchianti, G. Valtolina, J. A. Seman, E. Pace, M. De Pas, M. Inguscio, M. Zaccanti, and G. Roati, \textit{Phys. Rev. A} \textbf{90}, 043408 (2014).
\bibitem{Zurn2013} G. Z\"urn, T. Lompe, A. N. Wenz, S. Jochim, P. S. Julienne and J. M. Hutson, \textit{Phys. Rev. Lett.} \textbf{110}, 135301 (2013).
\bibitem{MIT_solitons} M. J. H. Ku, W. Ji, B. Mukherjee, E. Guardado-Sanchez, L. W. Cheuk, T. Yefsah, and M. W. Zwierlein, \textit{Phys. Rev. Lett.} \textbf{113}, 065301 (2014).
\bibitem{Brantut_2015} D. Husmann, S. Uchino, S. Krinner, M. Lebrat, T. Giamarchi, T. Esslinger, and J.-P. Brantut, \textit{Science} \textbf{350}, 1498 (2015).
\bibitem{Eckel2016} S. Eckel, J. G. Lee, F. Jendrzejewski, C. J. Lobb, G. K. Campbell, and  W. T. Hill, III, \textit{Phys. Rev. A} \textbf{93}, 063619 (2016).
\bibitem{Lee2013}  J. G. Lee, B. J. McIlvain, C. J. Lobb, and W. T. Hill, III, \textit{Sci. Rep.} \textbf{3}, 1034 (2013).
\bibitem{MITEoS} M. J. H. Ku, A. T. Sommer, L. W. Cheuk, M. W. Zwierlein, \textit{Science} \textbf{335}, 563 (2012).
\bibitem{Forbes_2014} M. McNeil Forbes and R. Sharma, \textit{Phys. Rev. A} \textbf{90}, 043638 (2014).
\bibitem{Gan11} S. Gandolfi, K. E. Schmidt, and J. Carlson, \textit{Phys. Rev. A} \textbf{83}, 041601(R) (2011).
\bibitem{piazza} F. Piazza, L. A. Collins, and A. Smerzi, \textit{New J. Phys.} \textbf{13}, 043008 (2011).
\bibitem{Zaremba} E. Zaremba, \textit{Phys. Rev. A}  \textbf{57}, 518 (1998).
\bibitem{Tosi} P. Capuzzi, P. Vignolo, F. Federici, and M. P. Tosi, \textit{Phys. Rev. A} \textbf{73}, 021603(R) (2006).
\bibitem{Spuntarelli} A. Spuntarelli, P. Pieri, and G. C. Strinati, \textit{Phys. Rev. Lett.} \textbf{99}, 040401 (2007).
\bibitem{PiazzaJOPB} F. Piazza, L. A. Collins, and A. Smerzi, \textit{J. Phys. B: At. Mol. Opt. Phys.}, \textbf{46}, 095302 (2013).
\bibitem{Abad} M. Abad, M. Guilleumas, R. Mayol, F. Piazza, D. M. Jezek, and A. Smerzi, \textit{Eur. Phys. Lett.} \textbf{109}, 40005 (2015).
\bibitem{Proukakis} K. Xhani, N. Proukakis \textit{et al.}, in preparation.
\end{thebibliography}
\end{document}